\documentclass[aoas]{imsart}

\RequirePackage{amsthm,amsmath,amsfonts,amssymb} \RequirePackage[authoryear]{natbib}
\RequirePackage[colorlinks,citecolor=blue,urlcolor=blue]{hyperref}
\RequirePackage{graphicx}
\usepackage{enumitem} \usepackage{bbm} \usepackage{color} \usepackage{xcolor}
\usepackage{subcaption} \usepackage{float} \usepackage{chngcntr} \usepackage{mathtools}
\usepackage{tikz} \usepackage{tcolorbox}
\usepackage{cleveref}
\usepackage{booktabs} \usepackage{multirow} \usepackage{booktabs}
\usetikzlibrary{tikzmark, decorations.pathreplacing, calc, positioning, bayesnet}

\startlocaldefs
\newtheorem{assumption}{Assumption}[section]
\crefname{assumption}{assumption}{Assumptions}

\endlocaldefs

\newcommand{\indep}{\rotatebox[origin=c]{90}{$\models$}}

\newcommand{\exposuredata}{\ensuremath{\hat{\theta}_{X_i}}}

\newcommand{\outcomedata}{\ensuremath{\hat{\theta}_{Y_i}}}

\newcommand{\exposuretrue}{\ensuremath{\theta_{X_i}}}

\newcommand{\outcometrue}{\ensuremath{\theta_{Y_i}}}

\newcommand{\causal}{\ensuremath{\beta}}

\newcommand{\params}{\ensuremath{\varphi}}

\newcommand{\p}{\ensuremath{p}}

\newcommand{\D}{\ensuremath{\mathbf{D}}}

\newcommand{\U}{\ensuremath{\mathbf{\Theta}}}

\newcommand{\completellk}{\ensuremath{l(\params; \U)}}

\newcommand{\clust}{\ensuremath{\xi}}

\usepackage{xcolor} \definecolor{qz}{RGB}{255,0,0}

\newcommand{\name}{MR-Path}

\begin{document}

\begin{frontmatter}
  \title{A Latent Mixture Model for Heterogeneous Causal Mechanisms in Mendelian
    Randomization}

  \begin{aug}
    \author[A]{\fnms{Daniel} \snm{Iong}\ead[label=e1,mark]{daniong@umich.edu}},
    \author[B]{\fnms{Qingyuan} \snm{Zhao}\ead[label=e2]{qyzhao@statslab.cam.ac.uk}} \and
    \author[A]{\fnms{Yang} \snm{Chen}\ead[label=e3]{ychenang@umich.edu}}
    \address[A]{Department of Statistics, University of Michigan, Ann Arbor,
      \printead{e1,e3}}

    \address[B]{Statistical Laboratory, University of Cambridge, \printead{e2}}
  \end{aug}

\begin{abstract}
  Mendelian Randomization (MR) is a popular method in epidemiology and genetics that
  uses genetic variation as instrumental variables for causal inference. Existing MR
  methods usually assume most genetic variants are valid instrumental variables that
  identify a common causal effect. There is a general lack of awareness that this effect
  homogeneity assumption can be violated when there are multiple causal pathways
  involved, even if all the instrumental variables are valid. In this article, we
  introduce a latent mixture model \name{} that groups instruments that yield similar
  causal effect estimates together. We develop a Monte-Carlo EM algorithm to fit this
  mixture model, derive approximate confidence intervals for uncertainty quantification,
  and adopt a modified Bayesian Information Criterion (BIC) for model selection. We
  verify the efficacy of the Monte-Carlo EM algorithm, confidence intervals, and model
  selection criterion using numerical simulations. We identify potential mechanistic
  heterogeneity when applying our method to estimate the effect of high-density
  lipoprotein cholesterol on coronary heart disease and the effect of adiposity on type
  II diabetes.
\end{abstract}


\end{frontmatter}


\section{Introduction} \label{sec:intro}

Mendelian randomization (MR) is a causal inference method that aims to estimate the
causal effect of a modifiable risk exposure on disease outcomes. MR is a special case of
instrumental variable methods that have a long history in statistics and econometrics.
The key insight of MR is that genetic variants, usually in the form of single nucleotide
polymorphisms (SNPs), are naturally randomised during conception and may serve as good
instrumental variables for many epidemiological risk factors
\citep{smith2004mendelian,Didelez2007}. As a study design, MR has been quickly gaining
popularity among epidemiologists because of its ability to give unbiased causal effect
estimates in the presence of unmeasured confounding and the increasing availability of
genome-wide association studies (GWAS) data.

A key assumption of MR is that the genetic instrumental variables can only affect the
outcome variables through the risk exposure under investigation. This is often referred
to as the ``exclusion restriction'' or ``no direct effect'' assumption in the
instrumental variable literature. With genetic variants as instruments, this assumption
may be violated due to a genetic phenomenon called ``pleiotropy'', meaning a single
genotype can affect multiple seemingly unrelated phenotypes. Recent empirical evidence
and genetic theory suggest that pleiotropy is pervasive for common traits
\citep{boyle2017expanded,liu2019trans}. This has lead to a burst of development of new
statistical methods aiming to make MR studies robust to different patterns of pleiotropy
\citep{bowden2015mendelian, kang2016instrumental, Zhao2018, verbanck2018detection,
  Burgess2020, Qi2019}.

To our knowledge, the vast majority of these robust MR methods still rely on the
``effect homogeneity'' assumption that the risk exposure has the same causal effect for
every individual. This assumption usually follows from assuming a linear structural
equation model commonly used in the instrumental variable literature
\citep{anderson1949estimation,bowden2017framework}. However, this key assumption may be
unrealistic when we use MR to study complex biological systems involving multiple
mechanisms, as demonstrated in the next
example. 

\subsection{Motivating example: The effect of HDL cholesterol on coronary heart disease}
\label{sec:motivating-example}

The statistical model we develop in this article is motivated by a real world problem.
Over the last few decades, there has been a heated debate in cardiology on the role of
high-density lipoproteins (HDL) in coronary heart disease (CHD)
\citep{rader2014hdl,daveysmith2020correlation}. Numerous observational studies have
found a consistent inverse association between HDL cholesterol (HDL-C, amount of
cholesterol carried in HDL particles) and CHD, lending support to a theory that HDL
plays a causally protective role in atherogenesis (formation of fatty deposits in the
arteries) through a biological mechanism called reverse cholesterol transport (HDL
particles remove excessive cholesterol in peripheral tissues). This led to a once widely
held belief among healthcare professionals and the general public that HDL particles are
the ``good cholesterol'', as opposed to low-density lipoproteins (LDL) which are thought
to be the ``bad cholesterol''.

However, the HDL hypothesis has received close scrutiny after several promising clinical
trials raising HDL cholesterol through the \emph{CETP} inhibitors demonstrated no or
only modest cardiovascular benefit \citep{armitage2019cholesteryl}. Moreover, a
prominent MR study further challenged the presumption that raising HDL-C will uniformly
translate into reductions in risk of CHD \citep{voight2012plasma}. Although there are
lots of SNPs associated with HDL-C, many of them are also associated with LDL
cholesterol and/or triglycerides. Due to this reason, \citet{voight2012plasma} based
their main argument on a SNP in the \emph{LIPG} gene that does not exhibit significant
association with LDL cholesterol and triglycerides, even though other genetic
instruments showed varied associations with risk of CHD. This shows that pleiotropy,
arising from the multiple mechanisms involved in the synthesis and regulation of blood
lipids, poses a major challenge for using MR methods to study HDL.

\begin{figure}
  \centering
  \begin{subfigure}{.5\textwidth}
    \centering \includegraphics[width=\columnwidth]{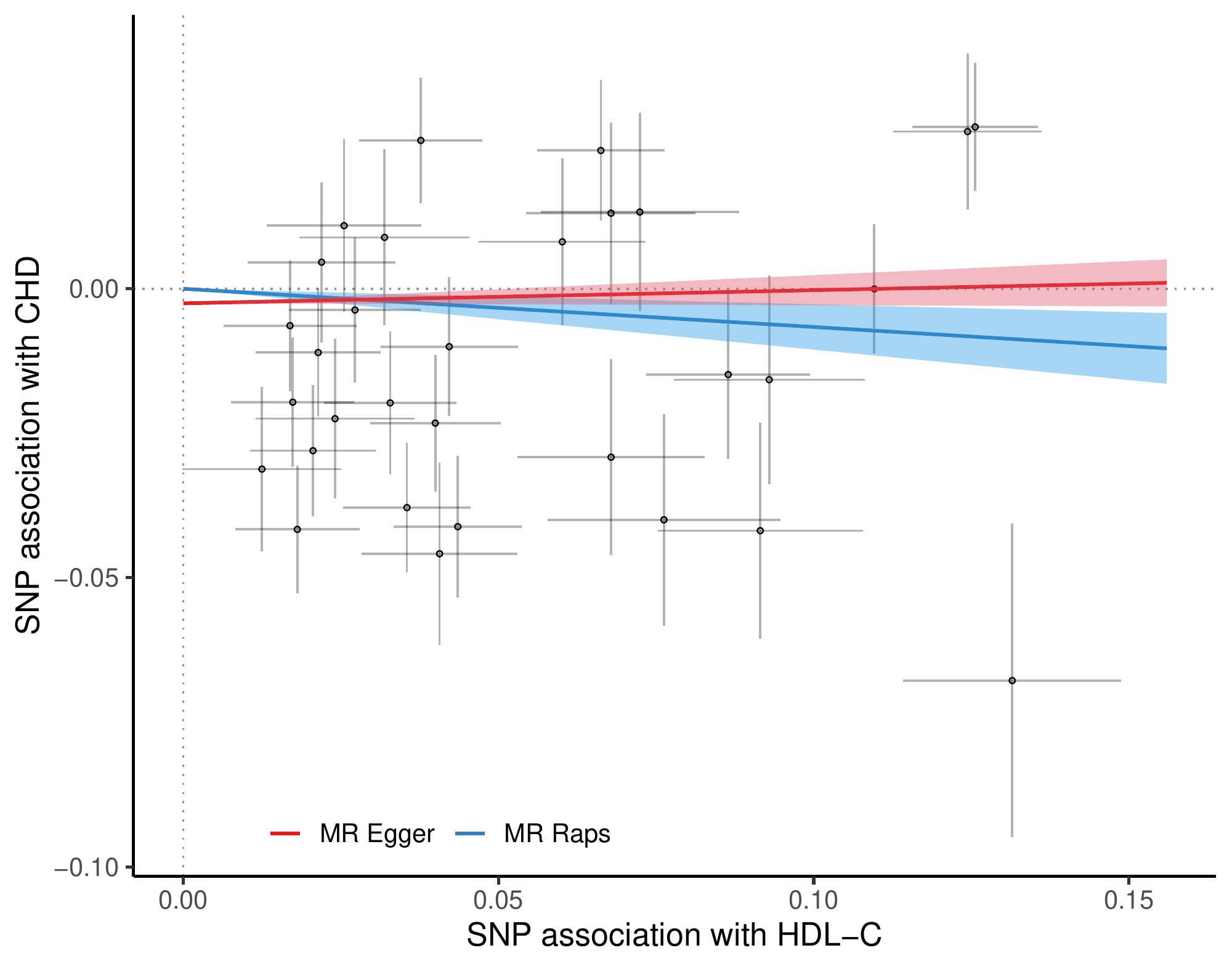}
    \caption{MR-RAPS \& MR-Egger}
    \label{fig:hdlcad-scatter-raps}
  \end{subfigure}%
  \begin{subfigure}{.5\textwidth}
    \centering \includegraphics[width=\columnwidth]{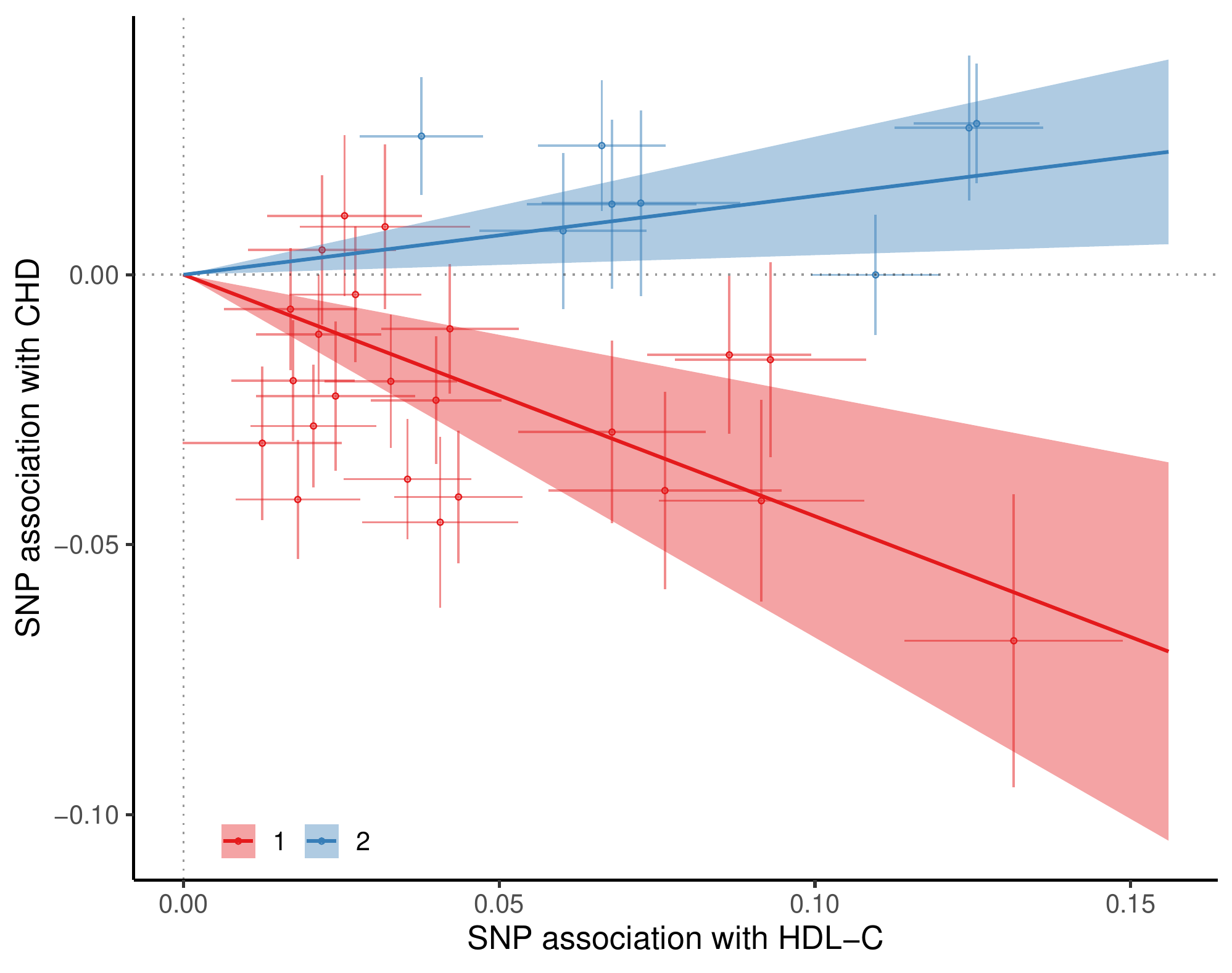}
    \caption{Proposed mixture model.}
    \label{fig:hdlcad-scatter-mixture}
  \end{subfigure}
  \caption{\textit{Scatterplot of HDL-CAD data and effect estimates}. \textbf{Left}: Line/shaded region
    represents the causal effect estimate $\pm$ one standard error from MR-RAPS.
    \textbf{Right}: Lines/shaded regions represent heterogeneous causal effect estimates
    $\pm$ one standard deviation from our proposed mixture model.}
  \label{fig:hdlcad-scatter}
\end{figure}

Using some of the latest large-scale GWAS datasets for HDL-C and CHD, we created a
dataset to visualize and analyze the heterogeneity among potential genetic instruments
for HDL (Figure \ref{fig:hdlcad-scatter}). Each point in this plot shows the reported
associations of a SNP with HDL-C and CHD in the GWAS (with standard error bars).
\Cref{fig:hdlcad-scatter-raps} shows straight lines across the origin whose slopes are
obtained using MR-RAPS \citep{Zhao2018} and MR-Egger \citep{bowden2015mendelian} which both assume a homogeneous effect of HDL-C on CHD.

If this assumption holds, the slopes in
\Cref{fig:hdlcad-scatter-raps} can be interpreted as the causal effects of HDL-C on CHD.
However, it is clear from \Cref{fig:hdlcad-scatter-raps} that the slopes estimated by
both MR-RAPS and MR-Egger provide a poor fit to the scatterplot.

In this paper, we propose to fit this dataset using an alternative model where the SNPs
have individual slopes that are drawn from a mixture distribution. This would be the
case if there are several biological mechanisms involved in regulating HDL-C; see
\Cref{sec:mech-het}. In this example, \name{} selects two clusters (shown in
\Cref{fig:hdlcad-scatter-mixture}) which provides a much better fit to the data. See
\Cref{sec:real-data} for more detail about the data collection for this example and the
results of our model.

\subsection{Related work and our contributions}
\label{sec:related-work}

There have been several attempts to develop MR methods that allow for heterogeneous
causal effects. We are also not the first to use mixture models for MR. The
\textit{contamination mixture} method proposed by \citet{Burgess2020} uses a
two-component mixture model to distinguish between valid and invalid instruments.
Similarly, the \textit{MR-Mix} method \citep{Qi2019} uses a four-component mixture model
to identify one group of valid instruments and three groups of invalid instruments which
have either direct effects on both the exposure and outcome, direct effects on the
outcome but no effect on the exposure, or no effect on both the exposure and outcome.
However, the purpose of using mixture models in these approaches is not to identify
different mechanisms but rather to provide a realistic model for the invalid
instruments. In particular, they assume that a plurality of the instruments are valid
and indicate an identical causal effect.

The only methods we are aware of that do not assume effect homogeneity and attempt to
distinguish causal mechanisms are \textit{GRAPPLE} \citep{Wang2020}, \textit{MR-Clust}
\citep{Foley2019} and \textit{BESIDE-MR} \citep{Shapland2020}. \textit{GRAPPLE} proposes
to use the local maximums of a robustified profile likelihood function \citep{Zhao2018}
to discover multiple mechanisms. However, \textit{GRAPPLE} is only a visualization tool
and does not attempt to explicitly model the different mechanisms. \textit{MR-Clust}
works by constructing a mixture model based on SNP-specific Wald estimators. Similar to
our proposed method, \textit{MR-Clust} does not make further assumptions about the
number of clusters and the structure of each cluster. However, a major limitation of
\textit{MR-Clust} is its assumption that the SNP-specific Wald estimates are normally
distributed, which is a poor approximation for weak instruments.
A comparison between our proposed method and
MR-Clust is provided in \cref{sec:comparison-with-mr} . \textit{BESIDE-MR} is another related method that uses Bayesian model averaging. Although \textit{BESIDE-MR}
is initially motivated by averaging over the uncertainty in selecting the valid
instruments, it can also be extended to allow for multiple clusters of instruments
indicating different causal effects. However, \textit{BESIDE-MR} is not a full
likelihood approach because it uses the profile likelihood derived in \citet{Zhao2018}
to eliminate the nuisance parameters related to the SNP-exposure effects.

Our paper makes two main contributions to this fast growing literature. First, there is
a general lack of awareness that MR can be used to discover multiple biological
mechanisms, partly due to the wide usage of the broad terminology ``effect
heterogeneity'' to refer to several different phenomena---invalid instrument due to
pleiotropy, effect modification/moderation by a covariate, and effect heterogeneity due
to different causal mechanisms. In this article we introduce the concept of mechanistic
heterogeneity for the last phenomenon and show that it can occur even if all the
instruments are valid.

Our second contribution is a transparent mixture model, which we call \name, to capture
the mechanistic heterogeneity. Because our model is based on the SNP-exposure and
SNP-outcome associations, it does not require the individual instruments to be strong.
We develop a Monte-Carlo EM algorithm to fit this model. Since our Monte Carlo EM
algorithm maximizes the full likelihood function for the latent mixture model, it has
all the benefits of likelihood-based inference.

The rest of the paper is organized as follows. In \cref{sec:iv}, we give a brief review
of the standard assumptions in MR. In \cref{sec:mech-het}, we introduce the concept of
mechanistic heterogeneity. In \cref{sec:mech-het-model} we propose to model it with
\name. In \cref{sec:inference}, we describe an Monte Carlo EM algorithm to fit \name{}
and discuss the relevant statistical inference and model selection procedures. We then
study the performance of our Monte Carlo EM algorithm with two simulation studies in
\cref{sec:sim} and apply it to two real data datasets (including the HDL-CHD example
above) in \cref{sec:real-data}. We conclude the paper with some ending remarks in
\cref{sec:discussion}.

\section{MR as an instrumental variables method} \label{sec:iv}

The goal of MR is to estimate the causal effect of a risk exposure variable (X) on a
disease outcome variable (Y). In particular, we may be interested in the causal effect
of HDL cholesterol (X) on the risk of coronary heart disease (Y). Regression analyses
between X and Y are typically biased by unobserved confounding variables $U$. MR uses
$p$ genetic variants $Z_1, \dots, Z_p$ as instrumental variables to obtain an unbiased
causal effect estimate of X on Y. A genetic variant $Z$ is said to be a \textit{valid
  instrument} for estimating the causal effect of $X$ on $Y$ if it satisfies the
following assumptions:

\begin{assumption}[Relevance] \label{as:relevance} It is associated with the risk
  exposure, i.e. $\text{Corr}(Z, X) \neq 0$.
\end{assumption}

\begin{assumption}[Independence] \label{as:independence} It must be independent of any
  unmeasured confounders that are associated with both the exposure and outcome, i.e.
  $Z \ \indep \ U$.
\end{assumption}

\begin{assumption}[Exclusion restriction] \label{as:excl-res} It affects the outcome
  only through the risk exposure, i.e. $Z \ \indep \ Y \ | \ X$.
\end{assumption}

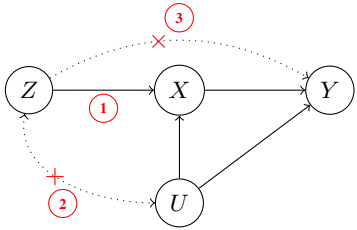
\begin{figure}
  \centering
  \begin{tikzpicture}
    \node[circle, draw] (X) at (0, 0) {$X$}; \node[circle, draw] (Y) at (2, 0) {$Y$};
    \node[circle, draw] (U) at (0, -1.5) {$U$}; \node[circle, draw] (Z) at (-2, 0)
    {$Z$}; \draw[->] (X) -- (Y); \draw[->] (U) -- (Y); \draw[->] (U) -- (X); \draw[->]
    (Z) to node[midway, draw, solid, font=\scriptsize, below, shape=circle, red,
    scale=0.7, yshift=-2] {\textbf{1}} (X); \draw[<->, dotted] (Z) to[out=-100,in=-180]
    node[midway, draw, solid, font=\scriptsize, below, shape=circle, yshift=-3, red,
    scale=0.7] {\textbf{2}} node[pos=0.45, red,rotate=-45] {\small \textbf{$\times$}}
    (U); \draw[->, dotted] (Z) to[out=30,in=150] node[midway, draw, solid,
    font=\scriptsize, above, shape=circle, yshift=3, red, scale=0.7] {\textbf{3}}
    node[pos=0.42, red] {\small \textbf{$\times$}} (Y);
  \end{tikzpicture}
  \caption{A directed acyclic graph (DAG) illustrating the core assumptions for a valid
    instrument.}
  \label{fig:iv-dag}
\end{figure}

\Cref{fig:iv-dag} provides a graphical representation of these assumptions. The
independence assumption is usually guaranteed by Mendel's law of random assortment of
genes. The relevance assumption is justified if genetic variants are chosen to be
genome-wide significant. Among the three assumptions, the exclusion restriction (ER)
assumption is the most problematic due to pleiotropy. Assessment of the IV assumptions
and sensitivity analysis in MR are discussed by \citet{Burgess2017}. In the MR
literature, it is common to assume the following linear structural equation model for
the exposure and outcome variables \citep{bowden2017framework}:
\begin{align}
  X & = \sum_{i=1}^p \exposuretrue Z_i + \eta_X U + E_X, \label{eq:linear-model-X}        \\
  Y & = \causal X + \sum_{i=1}^p \alpha_i Z_i + \eta_Y U + E_Y, \label{eq:linear-model-Y}
\end{align}
where $\eta_X$ and $\eta_Y$ are confounding effects and $E_X$ and $E_Y$ are random noise
terms acting on $X$ and $Y$ respectively. The causal effect between $X$ and $Y$ is given
by the parameter $\causal$. The true marginal association between $X$ and $Z_i$ is given
by $\exposuretrue$. The direct effect of $Z_i$ on $Y$ is given by $\alpha_i$.
\Cref{as:relevance} implies $\exposuretrue \neq 0$ and \cref{as:independence} implies
$Z_1,\dots,Z_p \ \indep \ U, E_X, E_Y$. Under this model, the exclusion restriction
assumption is violated if $\alpha_i \neq 0$. In particular, this can occur if $Z_i$
affects $Y$ through a mechanism unrelated to $X$ (see \cref{fig:multiple-direct-effect}
for illustration). Plugging \cref{eq:linear-model-X} into \cref{eq:linear-model-Y}, we
obtain
\begin{equation}
  Y = \sum_{i=1}^p \Big( \causal \exposuretrue + \alpha_i \Big) Z_i + (\eta_X + \eta_Y) U + (E_X + E_Y) = \sum_{i=1}^p \outcometrue Z_i + E_Y',
\end{equation}
where $\outcometrue = \beta \exposuretrue + \alpha_i$ gives the true marginal
association between $Y$ and $Z_i$ and $E_Y'$ is a random noise term that is independent
of $Z_i$. In summary-data MR, we usually observe the estimated SNP-exposure effect
$\exposuredata$, with standard error $\sigma_{X_i}$, and the estimated SNP-outcome
effect $\outcomedata$, with standard error $\sigma_{Y_i}$, for SNP $i = 1, \dots, p$.
These estimated effects are typically computed from two different samples using linear
or logistic regression. If we assume the genetic variants $Z_1, \dots, Z_p$ satisfy the
exclusion restriction assumption, in other words $\alpha_1 = \dots = \alpha_p = 0$, then
the causal effect parameter $\causal$ can be estimated consistently with the
inverse-variance-weighted estimator \citet{Burgess2013}. A more robust approach is to
perform error-in-variables regression of $\outcomedata$ on $\exposuredata$
\citep{Zhao2018}. We adopt this approach in \name. If the exclusion restriction
assumption is violated for some SNPs, then the causal effect parameter $\causal$ cannot
be identified without further assumptions on $\alpha_i$. For example, \citet{Zhao2018}
assumes that $\alpha_i \sim N(0, \tau^2)$ for most genetic variants so that the direct
effects are balanced out. 

\section{Mechanistic heterogeneity in MR}
\label{sec:mech-het}

\tikzset{SNP/.style = {draw, circle, minimum size=1.5cm}} \tikzset{M/.style = {draw,
    circle, minimum height=1.5cm, minimum width=1.5cm}} \tikzset{Xcomp/.style={draw,
    circle, minimum height=1.7cm, minimum width=1.7cm}} \tikzset{>=latex, shorten
  >=.5pt}


\begin{figure}
  \begin{subfigure}[t]{.45\linewidth}
    \centering
    \begin{tikzpicture}[every node/.style={scale=0.5}] \Large \def \firstcolor {red}
      \def \secondcolor {blue} \def \thirdcolor {violet}

      \node[SNP, \firstcolor] (Z1-1) {$Z_{1,1}$}; \node[\firstcolor, below of=Z1-1]
      (dots1) {\vdots}; \node[SNP, \firstcolor, below=.1cm of dots1.south] (Z1-p)
      {$Z_{1,p_1}$};

      \node[SNP, \secondcolor, below=.5cm of Z1-p] (Z2-1) {$Z_{2,1}$};
      \node[\secondcolor, below of=Z2-1] (dots2) {\vdots}; \node[SNP, \secondcolor,
      below=.1cm of dots2] (Z2-p) {$Z_{2,p_2}$};

      \node[SNP, \thirdcolor, below=.5cm of Z2-p] (Z3-1) {$Z_{3,1}$}; \node[\thirdcolor,
      below of=Z3-1] (dots3) {\vdots}; \node[SNP, \thirdcolor, below=.1cm of dots3]
      (Z3-p) {$Z_{3,p_3}$};

      \node[M, \firstcolor, right=1.2cm of dots1] (M1) {$M_1$};
      \draw[->, \firstcolor] (Z1-1) to (M1); \draw[->, \firstcolor] (Z1-p) to (M1);

      \node[M, \secondcolor, right=1.2cm of dots2] (M2) {$M_2$};
      \draw[->, \secondcolor] (Z2-1) to (M2); \draw[->, \secondcolor] (Z2-p) to (M2);

      \node[M, \thirdcolor, right=1.2cm of dots3] (M3) {$M_3$};
      \draw[->, \thirdcolor] (Z3-1) to (M3); \draw[->, \thirdcolor] (Z3-p) to (M3);

      \node[draw, circle, minimum height=1.7cm, minimum width=1.7cm, right=1.5cm of M2]
      (X) {$X$}; \draw[->, \firstcolor] (M1) to node[above] {$\theta_1$} (X); \draw[->,
      \secondcolor] (M2) to node[above] {$\theta_2$} (X); \draw[->, \thirdcolor] (M3) to
      node[above] {$\theta_3$} (X);

      \node[draw, circle, minimum height=1.7cm, minimum width=1.7cm, right=2 of X] (Y)
      {$Y$}; \draw[->] (X) to node[above] {$\beta$} (Y); \draw[->, \secondcolor, bend
      right] (M2) to node[below] {$\alpha_2$} (Y); \draw[->, \thirdcolor, bend right]
      (M3) to node[above] {$\alpha_3$} (Y);

      \node[draw, circle, minimum height=1.7cm, minimum width=1.7cm, above right=1cm of
      X] (U) {$U$}; \draw[->] (U) to (X); \draw[->] (U) to (Y);
    \end{tikzpicture}
    \caption{Scenario 1: Multiple pathways of horizontal pleiotropy.}
    \label{fig:multiple-direct-effect}
  \end{subfigure} %
  \hspace{35pt}
  \begin{subfigure}[t]{.45\linewidth}
    \centering
    \begin{tikzpicture}[every node/.style={scale=0.5}] \Large \def \firstcolor {red}
      \def \secondcolor {blue} \def \thirdcolor {violet}

      \node[SNP, \firstcolor] (Z1) {$Z_{1,1}$}; \node[\firstcolor, below of=Z1] (dots1)
      {\vdots}; \node[SNP, \firstcolor, below=.1cm of dots1.south] (Zj) {$Z_{1,p_1}$};

      \node[M, \firstcolor, right=1.2cm of dots1] (M1) {$M_1$};
      \draw[->, \firstcolor] (Z1) to (M1); \draw[->, \firstcolor] (Zj) to (M1);

      \node[Xcomp, \firstcolor, right=1cm of M1] (X1) {$X_1$}; \draw[->, \firstcolor]
      (M1) to node[above] {$\theta_1$} (X1);

      \node[SNP, \secondcolor, below=.5cm of Zj] (Zj1) {$Z_{2,1}$}; \node[\secondcolor,
      below of=Zj1] (dots2) {\vdots}; \node[SNP, \secondcolor, below=.1cm of dots2] (Zp)
      {$Z_{2,p_2}$};

      \node[M, \secondcolor, right=1.2cm of dots2] (M2) {$M_2$}; \draw[->, \secondcolor]
      (Zj1) to (M2); \draw[->, \secondcolor] (Zp) to (M2);

      \node[Xcomp, \secondcolor, right=1cm of M2] (X2) {$X_2$}; \draw[->, \secondcolor]
      (M2) to node[above] {$\theta_2$} (X2);

      \node[SNP, \thirdcolor, below=.5cm of Z2-p] (Z3-1) {$Z_{3,1}$}; \node[\thirdcolor,
      below of=Z3-1] (dots3) {\vdots}; \node[SNP, \thirdcolor, below=.1cm of dots3]
      (Z3-p) {$Z_{3,p_3}$};

      \node[M, \thirdcolor, right=1.2cm of dots3] (M3) {$M_3$};
      \draw[->, \thirdcolor] (Z3-1) to (M3); \draw[->, \thirdcolor] (Z3-p) to (M3);

      \node[Xcomp, \thirdcolor, right=1cm of M3] (X3) {$X_3$}; \draw[->, \thirdcolor]
      (M3) to node[above] {$\theta_3$} (X3);

      \node[draw, rectangle, minimum height=12.5cm, minimum width=2.7cm] (X) at
      ($(X1)!.5!(X3)$) {}; \node[below=.1cm of X.south] {\large $X = X_1 + X_2 + X_3$};

      \node[draw, circle, minimum height=1.7cm, minimum width=1.7cm, right=1.5cm of X2]
      (Y) {$Y$}; \draw[->, \firstcolor] (X1) to node[above] {$\beta_1$} (Y); \draw[->,
      \secondcolor] (X2) to node[above] {$\beta_2$} (Y); \draw[->, \thirdcolor] (X3) to
      node[above] {$\beta_3$} (Y);

      \node[draw, circle, minimum height=1.7cm, minimum width=1.7cm, above=1.5cm of Y]
      (U) {$U$}; \draw[->] (U) to (X1); \draw[->] (U) to (X2); \draw[->] (U) to (X3);
      \draw[->] (U) to (Y);

    \end{tikzpicture}
    \caption{Scenario 2: Multiple mechanisms for the exposure $X$.}
    \label{fig:multiple-x}
  \end{subfigure}
  \caption{Two scenarios of mechanistic heterogeneity}
  \label{fig:mechanistic-heterogeneity}
\end{figure}
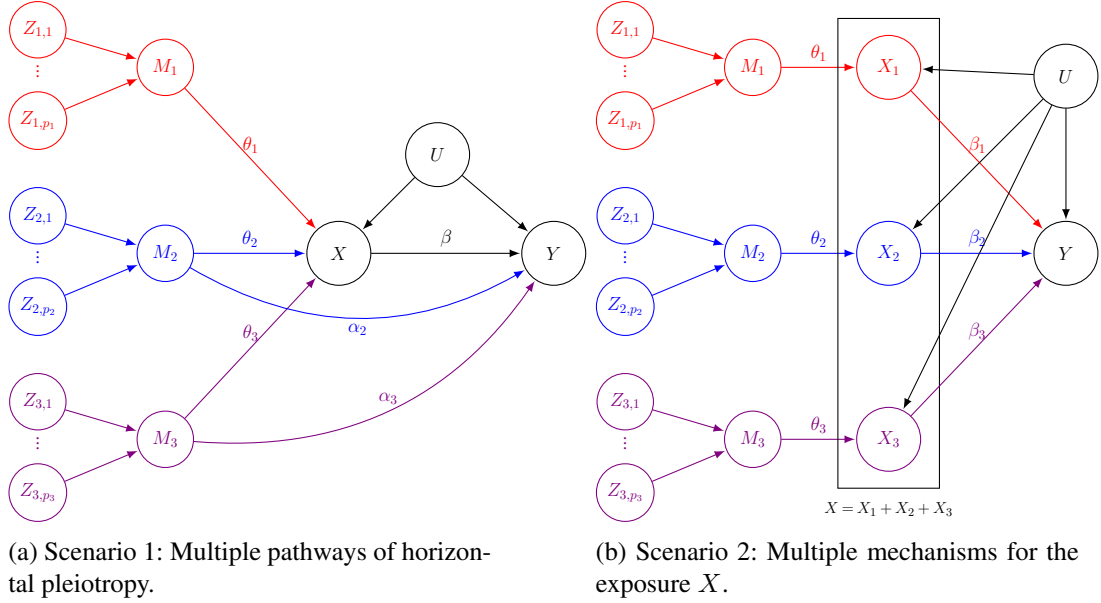

\begin{table}[h]
  \centering \small
  \caption{Ratio estimands using different instruments in the two scenarios in
    \Cref{fig:mechanistic-heterogeneity}. Directed acyclic graphs in
    \Cref{fig:mechanistic-heterogeneity} are interpreted as linear structural equation
    models.}
  \begin{tabular}{lcccc}
    \toprule
    {\bfseries Instruments $Z$} & {\bfseries Pathway $M$} & {\bfseries
                                                            Effect of
                                                            $M$ on
                                                            $X$}                        &
                                                                                          {\bfseries Effect of
                                                                                          $M$ on $Y$}                 &
                                                                                                                        {\bfseries
                                                                                                                        Wald estimand}
    \\
    \midrule
    \multicolumn{5}{c}{\bfseries Scenario 1}                                                  \\
    $Z_{1,1},\dotsc,Z_{1,p_1}$  & $M_1$                   & $\theta_1$ & $\theta_1 \beta$
                                                                                                                      &
                                                                                                                        $\beta$
    \\
    $Z_{2,1},\dotsc,Z_{2,p_2}$  & $M_2$                   & $\theta_2$ & $\theta_2 \beta
                                                                         + \alpha_2$                 &
                                                                                                       $\beta
                                                                                                       +
                                                                                                       \alpha_2/\theta_2$
    \\
    $Z_{3,1},\dotsc,Z_{3,p_3}$  & $M_3$                   & $\theta_3$ & $\theta_3 \beta
                                                                         + \alpha_3$                 &
                                                                                                       $\beta
                                                                                                       +
                                                                                                       \alpha_3/\theta_3$
    \\
    \midrule
    \multicolumn{5}{c}{\bfseries Scenario 2}                                                  \\
    $Z_{1,1},\dotsc,Z_{1,p_1}$  & $M_1$                   & $\theta_1$ & $\theta_1 \beta_1$
                                                                                                                      &
                                                                                                                        $\beta_1$
    \\
    $Z_{2,1},\dotsc,Z_{2,p_2}$  & $M_2$                   & $\theta_2$ & $\theta_2 \beta_2$ &
                                                                                              $\beta_2$
    \\
    $Z_{3,1},\dotsc,Z_{3,p_3}$  & $M_3$                   & $\theta_3$ & $\theta_3 \beta_3$ &
                                                                                              $\beta_3$                                                                                 \\
    \bottomrule
  \end{tabular}
  \label{tab:ratio-estimand}
\end{table}

The assumption that the direct effect $\alpha_i$ is iid normally distributed does not
take into account the possibility that genetic variation often affects phenotypic traits
through separate biological pathways. In this section we show that such behaviour may
lead to a clustering phenomenon where SNPs belonging to the same pathway would indicate
similar causal effects in an MR analysis. This is what we call ``mechanistic
heterogeneity'' in MR.

\subsection{Two origins of mechanistic heterogeneity}
\label{sec:origins}


Consider \Cref{fig:mechanistic-heterogeneity} which contains two scenarios of
mechanistic heterogeneity that motivate the latent mixture model. In both scenarios,
genetic variants are grouped into three biological pathways, $M_1$, $M_2$, and $M_3$,
that affect the exposure $X$ and outcome $Y$ differently. In the first scenario
(\Cref{fig:multiple-direct-effect}), all three pathways affect $X$ in the same way but
have different direct effects on $Y$. In particular, the first pathway $M_1$ does not
have any direct effect on $Y$ not mediated by $X$, so the instruments
$Z_{1,1},\dotsc,Z_{1,p_1}$ associated with it are all valid IVs.
In the second scenario in \Cref{fig:multiple-x}, the three pathways affect different
components of the exposure $X$ which also have different causal effects on the outcome
$Y$.
If we interpret the diagrams in \Cref{fig:mechanistic-heterogeneity} as linear
structural equations (like \Cref{fig:iv-dag} for \eqref{eq:linear-model-X} and
\eqref{eq:linear-model-Y}), we can derive the so-called Wald estimand (ratio of
$\theta_Y$ and $\theta_X$) for each instrument (\Cref{tab:ratio-estimand}). In both
scenarios in \Cref{fig:mechanistic-heterogeneity}, genetic instruments on the same
pathway have the same Wald estimand $\theta_Y/\theta_X$, while instruments across
different pathways generally have different estimands.

Therefore, we may reach completely different conclusions when using instruments on
different pathways in the MR analysis. In reality, mechanistic heterogeneity can be more
complicated than the basic scenarios in \Cref{fig:mechanistic-heterogeneity}. For
example, another pathway can affect some components of $X$ and also have direct effect
on $Y$. We study the robustness of our proposed method when horizontal pleiotropy and
multiple mechanisms are simultaneously present in \cref{sec:robustn-under-plei}.
It is also worthwhile to point out that mechanistic heterogeneity can arise even
when all the IVs are perfectly valid; an example is Scenario 2 in
\Cref{fig:mechanistic-heterogeneity}.

\subsection{Relationship to local average treatment effect}
\label{sec:relat-local-aver}

The above introduction of mechanistic heterogeneity is entirely based on linear
structural equation models. Next we show that the same clustering phenomenon can also
happen when there is nonlinearity. Consider the causal diagram in
\Cref{fig:multiple-direct-effect} without the $M_2 \to Y$ and $M_3 \to Y$ edges, so all
the instruments $Z_{1,1},\dotsc,Z_{3,p_3}$ are valid (satisfy
\Cref{as:relevance,as:independence,as:excl-res}). Suppose the variables satisfy the
nonparametric structural equation model (NPSEM) with independent errors
\citep{pearl2009causality} according to \Cref{fig:multiple-direct-effect}, so the
counterfactuals of $M_k, k =1,2,3$, $X$, and $Y$ can be defined using the NPSEM. For
example, we use $Y(X=1)$ to denote the counterfactual outcome under the intervention
$X=1$. To simplify our illustration below, we assume $Z_{1,1},\dotsc,Z_{3,p_3}$,
$M_1,M_2,M_3$, and $X$ are all binary variables.

It is well known that if the instrument $Z_{k,j}$ is valid and the counterfactuals of
the exposure $X$ satisfy the monotonicity assumption
$X(Z_{k,j} = 1) \ge X(Z_{k,j} = 0)$, the Wald estimand for instrument $Z_{k,j}$ is equal
to the so-called local average treatment effect,
$\mathbb{E}[Y(X=1) - Y(X=0) \mid X(Z_{k,j} = 1) > X(Z_{k,j} = 0)]$
\citep{angrist1996identification}. Notice that this interpretation of the IV analysis
does not require linearity of the structural equation model. Suppose we further assume
the effect of $Z$ on $M$ is monotone, $M_k(Z_{k,j} = 1) \ge M_k(Z_{k,j} = 0),$ and the
effect of $M$ on $X$ is monotone, $X(M_k = 1) \ge X(M_k = 0)$. Using the properties of
counterfactuals and the fact in \Cref{fig:multiple-direct-effect} that $Z_{k,j}$ affects
$X$ entirely through $M_k$, we get
\[
  X(Z_{k,j} = z) = X(Z_{k,j} = z, M_k = M_k(Z_{k,j} = z)) = X(M_k = M_k(Z_{k,j} = z)).
\]
Thus, using the assumption that $X$ and $M_k$ are binary,
\begin{equation} \label{eq:late-within-mechanism}
  \begin{split}
    &\mathbb{E}\big[Y(X=1) - Y(X=0) \mid X(Z_{k,j} = 1) > X(Z_{k,j} = 0)\big] \\
    =& \mathbb{E}\big[Y(X=1) - Y(X=0) \mid X\big(M_k = M_k(Z_{k,j} = 1)\big) >
    X\big(M_k = M_k(Z_{k,j} = 0)\big)\big] \\
    =& \mathbb{E}\big[Y(X=1) - Y(X=0) \mid X(M_k = 1) >
    X(M_k = 0), M_k(Z_{k,j} = 1) > M_k(Z_{k,j} = 0) \big] \\
    =& \mathbb{E}\big[Y(X=1) - Y(X=0) \mid X(M_k = 1) > X(M_k = 0) \big].
  \end{split}
\end{equation}
The last equality above uses
\[
  \{M_k(Z_{k,j} = 0), M_k(Z_{k,j} = 1)\}\, \indep \,\{X(M_k = 0), X(M_k = 1), Y(X = 0), Y(X = 1)\}.
\]
This counterfactual independence follows from expressing the counterfactuals using the
NPSEM and the assumption that the different structural equations have independent
errors.

The significance of \eqref{eq:late-within-mechanism} is that, if the counterfactuals of
$M$ and $X$ satisfy the monotonicity assumption, the local average treatment effect
corresponding to $Z_{k,j}$ only depends on the mechanism index $k$. This shows that the
clustering of the Wald esimand in \Cref{sec:origins} not only occurs in linear
structural equation models but also in certain nonlinear models. These examples
demonstrate the importance of identifying mechanistic heterogeneity to correctly
interpret MR studies.

\section{\name: A latent mixture model for mechanistic heterogeneity}
\label{sec:mech-het-model}

\begin{figure}[ht]
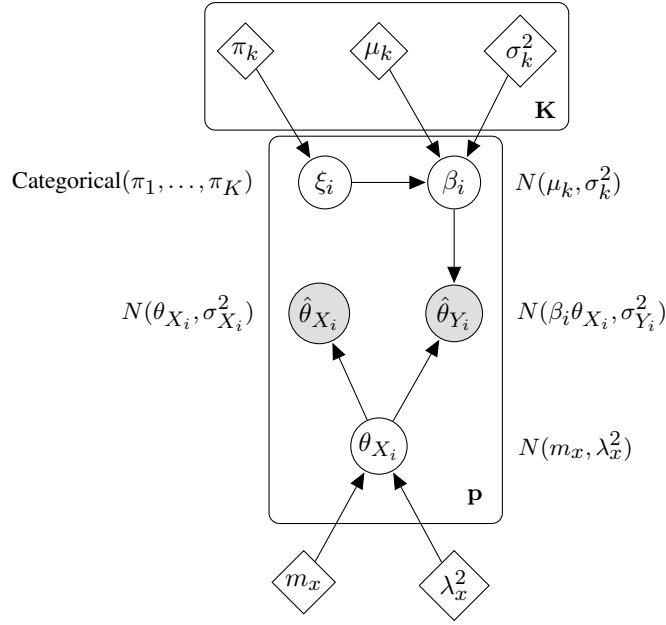

  \centering \tikz{ \node[obs] (exposuredata) {$\exposuredata$}; \node[obs, right=of
    exposuredata] (outcomedata) {$\outcomedata$}; \node[latent, above=of outcomedata]
    (beta) {$\causal_{i}$}; \node[latent, left=of beta] (clust) {$\clust_{i}$};
    \node[latent, below=of outcomedata, xshift=-1cm] (exposuretrue) {$\exposuretrue$};
    \node[det, above=of beta, xshift=-1cm] (mu_k) {$\mu_{k}$}; \node[det, right=of mu_k]
    (sigma_k) {$\sigma_{k}^{2}$}; \node[det, left=of mu_k] (pi_k) {$\pi_{k}$};
    \node[det, below=of exposuretrue, xshift=-1cm] (m_x) {$m_{x}$}; \node[det, below=of
    exposuretrue, xshift=1cm] (lambda_x) {$\lambda_{x}^{2}$};

    \edge {pi_k} {clust}; \edge {mu_k} {beta}; \edge {sigma_k} {beta}; \edge{clust}
    {beta}; \edge {beta} {outcomedata}; \edge {exposuretrue} {exposuredata}; \edge
    {exposuretrue} {outcomedata}; \edge {m_x} {exposuretrue}; \edge {lambda_x}
    {exposuretrue};

    \plate[inner sep=0.25cm] {plate1} {(beta) (clust) (exposuredata) (outcomedata)
      (exposuretrue)} {$\mathbf{p}$}; \plate[inner sep=0.15cm] {plate2} {(pi_k) (mu_k)
      (sigma_k)} {$\mathbf{K}$};

    \node[right=of beta, xshift=-19pt] (beta_as) {$N(\mu_{k}, \sigma_{k}^{2})$};
    \node[left=of clust, xshift=15pt] (clust_as)
    {$\text{Categorical}(\pi_{1},\dots,\pi_{K})$}; \node[right=of exposuretrue,
    xshift=10pt] (exposuretrue_as) {$N(m_{x}, \lambda_{x}^{2})$}; \node[right=of
    outcomedata, xshift=-19pt] (outcomedata_as)
    {$N(\beta_{i} \exposuretrue, \sigma_{Y_{i}}^{2})$}; \node[left=of exposuredata,
    xshift=19pt] (exposuredata_as) {$N(\exposuretrue, \sigma_{X_{i}}^{2})$}; }
  \caption{Graphical model formulation of MR-Path. Observed data is represented by gray
    circles; latent variables are represented by white circles; and model parameters are
    represented by diamonds.}
  \label{fig:graph-model}
\end{figure}

Motivated by the observations in the previous section, we propose a latent mixture model
to discover mechanistic heterogeneity using summary GWAS data. In essence, this model
assumes that each genetic variant has a specific causal effect and the genetic variants
on the same biological pathway have similar variant-specific causal effects and form
clusters. The mean of each cluster corresponds to the Wald estimand of that pathway
(last column in \cref{tab:ratio-estimand}). These assumptions, along with standard
assumptions for summary-data MR literature, are introduced below. A graphical model
formulation of MR-Path is shown in \cref{fig:graph-model}.

%
%

\begin{assumption}[Error-in-variables regression] The observed instrument-exposure and
  instrument-outcome associations are distributed as
  \begin{equation}
    \begin{pmatrix} \exposuredata \\ \outcomedata \end{pmatrix} \overset{\text{indep.}}{\sim} N \Big( \begin{pmatrix} \exposuretrue \\ \causal_i \exposuretrue \end{pmatrix}, \begin{pmatrix} \sigma_{X_i}^2 & 0 \\ 0 & \sigma_{Y_i}^2 \end{pmatrix} \Big), \quad i = 1,\dots, p, \label{eq:eiv-ext}
  \end{equation}
  where $\sigma_{X_i}$, $\sigma_{Y_i}$ are (fixed) measurement errors.
  \label{as:data-model}
\end{assumption}

In this assumption, the variant-specific causal effects are given by $\causal_i$. The
normality assumption is justified because $\exposuredata$ and $\outcomedata$ are
typically linear (or logistic) regression coefficients which are computed in GWAS with a
large sample size. Independence between $\exposuredata$ and $\outcomedata$ for each SNP
is justified if they are from GWAS conducted with non-overlapping samples. Independence
of the estimated effects across different SNPs is a reasonable assumption if we select
SNPs that are uncorrelated by using linkage disequilibrium clumping. Although
independence between SNPs does not imply the estimated effects are uncorrelated, the
correlation between the estimated effects are typically negligible \citep{Zhao2018}.

\begin{assumption}[Mixture model for mechanistic heterogeneity] \label{as:mixture-model}
  \begin{align}
    \clust_i                 & \sim \text{Categorical }(\pi_1,\dots,\pi_K), \label{eq:Z_i}                \\
    \causal_i | \clust_{i} = k & \sim N(\mu_k, \sigma_k^2), \quad k = 1,\ldots, K. \label{eq:mixture-model}
  \end{align}
\end{assumption}

We assume a Gaussian mixture model for the variant-specific (latent) causal effects
$\causal_i$ to probabilistically cluster genetic variants with similar causal effects.
An indicator variable for cluster membership of SNP $i$ is given by $\clust_{i}$. We can
compute the posterior distribution of $\clust_{i}$ and $\causal_i$ to summarize our
knowledge of these variant-specific latent variables based on data (see
\cref{sec:post-sampling}). The $K$ clusters represent different causal mechanisms where
the cluster means $\mu_k$, for $k = 1,\dots,K$, identify the average causal effects for
each mechanism. The cluster proportions $\pi_k$ give us the proportion of genetic
variants associated with each mechanism. The cluster variances $\sigma_k^2$ quantify
uncertainty within each mechanism. Therefore, the parameters of interest in \name{} are
given by $\params = \{\pi_k, \mu_k, \sigma_{k}^{2}: k = 1,\dots,K\}$.
Note that \cref{eq:mixture-model} assumes the Wald estimand corresponding to each
genetic instrument $\beta_{i}$ is allowed to fluctuate around the cluster means
$\mu_{k}$'s. The cluster variance $\sigma_{k}^{2}$ thus captures the within-cluster
heterogeneity due to unaccounted direct effects of the instruments on the outcome. This
is similar to making the InSIDE (INstrument Strength Independent of Direct Effect)
assumption \citep{bowden2015mendelian} within each mixture component.
The number of clusters $K$ is unknown but can be chosen based on heuristics and domain
knowledge or estimated from data---see \cref{sec:model-selection} for a model selection
criterion for choosing $K$. Note that \cref{as:data-model} implies $\causal_{i}$ and
$\exposuretrue$ are marginally independent but conditionally dependent given the
observed data. This can be deduced from the graphical model in \cref{fig:graph-model}
(chapter 8 in \cite{bishop2006pattern}).

\section{Statistical inference for \name} \label{sec:inference}

In order to fit \name{} to gain insight into mechanistic heterogeneity, we proceed by
discussing three inference procedures. First, we describe our implementation of the EM
algorithm for obtaining a maximum likelihood estimate of $\params$ and point out the
challenges and our solutions for the expectation step. Second, we discuss approximate
confidence intervals for $\params$ obtained by computing and inverting the observed
information matrix. Lastly, we go into how $K$ can be selected from data using a
modified Bayesian Information criterion (BIC). Let
$\D = \{(\exposuredata, \sigma_{X_i}, \outcomedata, \sigma_{Y_i}): i =1,\dots,\p \}$
denote the observed data and let
$\U = \{(\exposuretrue, \causal_i, \clust_i): i = 1,\dots,\p\}$ denote the set of latent
variables.

\subsection{Overview of the EM Algorithm}

The Expectation-Maximization (EM) algorithm is an iterative procedure commonly used to
perform maximum likelihood estimation in latent variable models. To define the EM
algorithm for \name, we derive two important model quantities: the complete-data log
likelihood and the conditional posterior of the latent variables $\U$, given parameters
$\params$. The latter is used to compute the Q-function in the expectation step. Let
$\phi(\cdot; \mu, \sigma^2)$ denote the density of $N(\mu, \sigma^2)$. We make the
additional model assumption that $\exposuretrue \sim N(\nu_x, \lambda_x^2)$. It follows
that the complete-data log likelihood for \name{} is given by
\begin{align}
  \completellk := \sum_{i=1}^\p l_i(\params; \exposuretrue, \causal_i, \clust_i), \label{eq:completeDataLogLik}
\end{align}
where
\begin{equation}
  l_i(\params; \exposuretrue, \causal_i, \clust_i) \propto \log \phi(\exposuretrue; \nu_x, \lambda_x^2) + \sum_{k=1}^K Z_{ik} \big[ \log \pi_k + \log \phi(\causal_i; \mu_k, \sigma_k^2) \big] \label{eq:l-i}
\end{equation}
and $Z_{ik} = 1$, if $\clust_i = k$, and $0$ otherwise. The conditional posterior of the
latent variables given $\params$ can be decomposed as
\begin{align}
  P(\U| \D, \params) := & \prod_{i=1}^\p P(\causal_i, \clust_i, \exposuretrue | \exposuredata, \outcomedata, \params) \nonumber                                                                                                                        \\
  =                     & \prod_{i=1}^\p \prod_{k=1}^K \big[P(\clust_i = k |\causal_i, \params)\big]^{Z_{ik}} P(\causal_i | \exposuretrue, \outcomedata, \params) P(\exposuretrue | \exposuredata, \outcomedata, \params), \label{eq:latent-posterior}
\end{align}
where $P(\clust_i = k | \causal_i, \params)$ and
$P(\causal_i | \exposuretrue, \outcomedata, \params)$ are available in closed-form and
are given by
\begin{align}
  P(\clust_i = k | \causal_i, \params)                     & = \frac{\pi_k \phi(\causal_i; \mu_k, \sigma_k^2)}{\sum_{j=1}^K \pi_j \phi(\causal_i; \mu_j, \sigma_j^2)} \label{eq:Zi-dist}, \\
  P(\causal_i | \exposuretrue, \outcomedata, \params) & = \sum_{k=1}^K \tilde{\pi}_{ik} \phi(\causal_i; \tilde{\mu}_{ik}, \tilde{\sigma}_{ik}^2); \label{eq:beta-dist}
\end{align}
where
\begin{equation}
  \tilde{\pi}_{ik} = \pi_{k} \Big[ 2\pi \Big(\frac{\exposuretrue^{2}}{\sigma_{Y_{i}}^{2}} + \frac{1}{\sigma_{k}^{2}}  \Big)^{-1} \Big]^{1/2},\quad
  \tilde{\sigma}_{ik}^2 = \Big(\frac{1}{\sigma_k^2} + \frac{\exposuretrue^2}{\sigma_{Y_i}^2}\Big)^{-1},\quad    \tilde{\mu}_{ik}  = \tilde{\sigma}_{ik}^2 \Big( \frac{\outcomedata \exposuretrue}{\sigma_{Y_i}^2} + \frac{\mu_k}{\sigma_k^2} \Big).
  \label{eq:beta-dist-params}
\end{equation}

Details for the derivation of $P(\causal_{i} | \exposuretrue, \outcomedata, \params)$
are given in \cref{sec:derivation}.
Unfortunately, instead of having an analytical solution, the probability density
$P(\exposuretrue | \exposuredata, \outcomedata, \params)$ is known only up to a
multiplicative constant given by
\begin{align}
  P(\exposuretrue | \exposuredata, \outcomedata, \params) & \propto P(\exposuretrue | \exposuredata, \params) P(\outcomedata | \exposuretrue, \params) \nonumber                                                                             \\
                                                          & = \phi(\exposuretrue; m_{X_i}, \lambda_{X_i}^2) \sum_{k=1}^K \pi_k \phi(\outcomedata; \exposuretrue \mu_k, \exposuretrue^2 \sigma_k^2 + \sigma_{Y_i}^2) \label{eq:target-distr},
\end{align}
where
\begin{align*}
  \lambda_{X_i}^2 & = \Big( \frac{1}{\sigma_{X_i}^2} + \frac{1}{\lambda_x^2} \Big)^{-1},\quad
                    m_{X_i} = \lambda_{X_i}^2 \Big( \frac{\exposuredata}{\sigma_{X_i}^2} + \frac{\nu_x}{\lambda_x^2} \Big).
\end{align*}

The EM algorithm starts with initial values for the parameter estimates $\params^{(0)}$.
Each iteration $t = 1,2,\dots$ of the EM algorithm consists of an expectation step
(E-step) and a maximization (M-step). The E-step involves computing the Q-function,
defined as the expectation of the complete-data log likelihood with respect to the
conditional posterior of the latent variables given the previous iterations parameter
estimates. Specifically, the E-step can be represented by
\begin{equation}
  Q(\params, \params^{(t-1)}) = E \big[l(\params; \U) | \D, \params^{(t-1)} \big] \label{eq:Q-function},
\end{equation}

where the expectation is taken with respect to $P(\U | \D, \params)$. The M-step
consists of computing an update of the parameter estimates for the current iteration as
the value that maximizes the Q-function. In other words, the M-step involves
\begin{equation}
  \params^{(t)} = \underset{\params}{\text{arg max }} Q(\params, \params^{(t-1)}).
\end{equation}

The EM algorithm guarantees the likelihood is non-decreasing across iterations---ascent
property---and is theoretically guaranteed, under mild regularity conditions, to
converge to a local optimum \citep{Wu1983}. In practice, a commonly used heuristic for
determining whether the EM algorithm has converged is when the increase in the
Q-function from the previous iteration is less than a specified threshold.
Unfortunately, since $P(\U | \D,\params)$ is only known up to a multiplicative constant,
the Q-function for \name{} cannot be computed analytically. In the next section, we
discuss our implementation of a variant of the EM algorithm where the Q-function in the
E-step is approximated using Monte-Carlo methods---the Monte-Carlo EM (MC-EM) algorithm.


\subsection{Implementation of the Monte-Carlo EM Algorithm}

The MC-EM algorithm allows us to perform maximum likelihood estimation in latent
variable models where the Q-function cannot be computed analytically but can be
approximated using Monte-Carlo methods. In \cref{sec:MC-E-step}, we describe an
importance sampling (IS) scheme for approximating the Q-function in \name. The main
drawback of the MC-EM algorithm is that Monte-Carlo error accrued from approximating the
Q-function may cause the algorithm to not converge \citep{Neath2013}. More precise
approximations of the Q-function are needed as the M-step updates approach a local
optimum. Furthermore, the MC-EM algorithm does not satisfy the ascent property of the
vanilla EM algorithm. A simple solution to these issues is to set the number of
Monte-Carlo samples used to approximate the Q-function (MC sample size) to be very large
or increase the MC sample size by a deterministic large amount at each iteration.
However, this may not be computationally feasible, thus creating a trade-off between
statistical consistency and computational efficiency. In \cref{sec:ascent-mcem-section},
we discuss an automated data-driven procedure introduced by \citet{Caffo2005} that
guarantees the ascent property is satisfied with high probability by assessing
Monte-Carlo error at the end of each iteration and increasing the MC sample size
accordingly. The runtimes for the proposed MC-EM algorithm applied to the two examples in \cref{sec:real-data} are provided in \cref{sec:comp-efficiency}.


\subsubsection{Monte-Carlo E-step using Importance Sampling} \label{sec:MC-E-step}

The decomposition of $P(\U | \D, \params)$ in \cref{eq:latent-posterior} suggests that
we can obtain importance samples from $P(\U | \D, \params)$ by first drawing samples of
$\{\exposuretrue\}$ from an importance (proposal) distribution. Given importance samples
of $\{\exposuretrue\}$, we can obtain importance samples of $\{\causal_i\}$ and
$\{\clust_i\}$ by directly sampling from
$\prod_{i=1}^p P(\causal_i | \exposuretrue, \exposuredata, \params)$ and
$\prod_{i=1}^p P(\clust_i | \causal_i, \params)$, which are available in closed form in
\cref{eq:Zi-dist} and \cref{eq:beta-dist}. Then, we can estimate the Q-function with a
sum of the complete-data log-likelihood evaluated at the samples weighted by the
importance weights.

Choosing an importance distribution that yields an efficient sampling procedure is a
non-trivial task that depends on the form of the target distribution (see
\citet{Tokdar2010} for a review). In our case, the target distribution is
$\prod_{i=1}^p P(\exposuretrue | \exposuredata, \outcomedata, \params)$ so a sensible
choice for an importance (proposal) distribution would be
$\prod_{i=1}^p P(\exposuretrue | \exposuredata, \params)$, the posterior distribution of
$\{\exposuretrue\}$ given only $\{\exposuredata\}$ instead of the full data
$\{\exposuredata, \outcomedata\}$ which has a closed form given in
\cref{eq:target-distr}. This choice of importance distribution yields importance weights
that are bounded and have a finite variance.

More precisely, let $m_t$ denote the number of desired importance samples at iteration
$t$ and $\tilde{\params}^{(t-1)}$ denote the MC-EM update from iteration $t-1$. For each
$i = 1,\dots,p$ and $j = 1,\dots, m_t$, suppose
$\exposuretrue^j \sim P(\exposuretrue | \exposuredata, \params^{(t-1)})$,
$\causal_i^j \sim P(\causal_i | \exposuretrue, \outcomedata, \params^{(t-1)})$, and
$\clust_i^j \sim P(\clust_i | \causal_i, \params^{(t-1)})$. From \cref{eq:target-distr},
we have that the unnormalized importance weights are given by
\begin{equation}
  w_i^j = P(\outcomedata | \exposuretrue^j, \tilde{\params}^{(t-1)}) = \sum_{k=1}^K \tilde{\pi}_k^{(t-1)} \phi \Big(\outcomedata;  \exposuretrue^j \tilde{\mu}_k^{(t-1)}, (\exposuretrue^j)^2 \tilde{\sigma}_k^{2 (t-1)} + \sigma_{Y_i}^2 \Big)
  \label{eq:impt-weights}
\end{equation}
for $j = 1,\dots, M$. It can be shown that
$0 \leq w_i^j \leq \big(2\pi \sigma_{Y_i}^2\big)^{-1/2}$ (proof is provided in
\cref{sec:IS-weights-bound}). Let $\bar{w}_i^j = w_i^j / \sum_{j=1}^{m_t} w_i^j$ be the
normalized importance weights. The IS estimate of the Q-function at iteration $t$ is
given by

\begin{equation}
  \tilde{Q}(\params, \tilde{\params}^{(t-1)}; m_t) = \sum_{i=1}^p \sum_{j=1}^{m_t} \bar{w}_i^j l_i^j(\params), \label{eq:tilde-Q}
\end{equation}
where
\begin{equation*}
  l_i^j(\params) = l_i(\params; \exposuretrue^j, \causal_i^j, \clust_i^j),
\end{equation*}
and $l_i$ is defined in \cref{eq:l-i}. Consequently, the MC-EM update of the model
parameters at iteration $t$ approximated with Monte-Carlo sample size $m_t$ is given by
\begin{equation*}
  \tilde{\params}^{(t, m_t)} = \underset{\params}{\text{arg max }} \tilde{Q}(\params, \tilde{\params}^{(t-1)}; m_t).
\end{equation*}

\subsubsection{Ascent-Based Monte-Carlo EM} \label{sec:ascent-mcem-section}

At the end of each MC-EM iteration, the ascent-based MC-EM algorithm~\citep{Caffo2005}
performs a hypothesis test to determine whether the Q-function has increased from the
previous iteration. If there is not sufficient evidence that suggests the Q-function has
increased, we reject the current iteration's M-step and repeat the iteration with a
larger MC sample size. In this section, we will describe this procedure concretely for
our inference problem.

Define the change in the Q function at iteration $t$ of MC-EM as
\begin{equation}
  \Delta Q(\tilde{\params}^{(t, m_t)}, \tilde{\params}^{(t-1)}) := Q(\tilde{\params}^{(t, m_t)}, \tilde{\params}^{(t-1)}) - Q(\tilde{\params}^{(t-1)}, \tilde{\params}^{(t-1)}). \label{eq:delta-Q}
\end{equation}

After obtaining $\tilde{\params}^{(t, m_t)}$ in the M-step, we are interested in
performing the following hypothesis test at a specified significance level $\alpha$:
\begin{align}
  H_0: \Delta Q(\tilde{\params}^{(t,m_t)}, \tilde{\params}^{(t-1)}) = 0, \label{eq:test1} \\
  H_A: \Delta Q(\tilde{\params}^{(t,m_t)}, \tilde{\params}^{(t-1)}) > 0. \nonumber
\end{align}

If we reject $H_0$, then we accept $\tilde{\params}^{(t, m_t)}$ and move on to the next
iteration. If we fail to reject $H_0$, then we reject $\tilde{\params}^{(t, m_t)}$ and
repeat the current iteration with a larger Monte-Carlo sample size. The change in the Q
function in \cref{eq:delta-Q} can be approximated using
\begin{equation}
  \Delta \tilde{Q}(\tilde{\params}^{(t, m_t)}, \tilde{\params}^{(t-1)}) \approx \sum_{i=1}^\p \sum_{j=1}^{m_t} w_i^j \Lambda_{ij}^{(t)} \label{eq:deltaQ-tilde}
\end{equation}

where
\begin{equation}
  \Lambda_{ij}^{(t)} = l_i^j(\tilde{\params}^{(t, m_t)})  - l_i^j(\tilde{\params}^{(t-1)}). \label{eq:lambda-ij}
\end{equation}

It was shown in \citet{Caffo2005} that under $H_0$,
\begin{equation}
  \sqrt{m_t}\Delta \tilde{Q}(\tilde{\params}^{(t,m_t)}, \tilde{\params}^{(t-1)}) \overset{d}{\rightarrow} N(0, \eta^2) \label{eq:deltaQ-asymp-dist}
\end{equation}
as $m_t \rightarrow \infty$, where $\sigma^2$ depends on the sampling procedure used. We
can obtain an estimate of $\eta^2$, $\hat{\eta}^2$, by computing the variance of the
importance sampling estimate in \cref{eq:deltaQ-tilde} (details provided in
\cref{sec:IS-eta}). Therefore, we can reject $H_0$ with approximate significance level
$\alpha$ if
\begin{equation*}
  \Delta \tilde{Q}(\tilde{\params}^{(t,m_t)}, \tilde{\params}^{(t-1)}) - z_{\alpha} \frac{\hat{\eta}}{m_t} > 0
\end{equation*}
where $z_{\alpha}$ is the $(1 - \alpha)^{\text{th}}$ quantile of the standard normal
distribution. If we fail to reject $H_0$, we repeat iteration $t$ with a larger $m_t$
until we are able to reject $H_0$. Similarly, a convenient stopping criterion is
obtained by testing
\begin{equation*}
  H_A: \Delta Q(\tilde{\params}^{(t,m_t)}, \tilde{\params}^{(t-1)}) < \epsilon
\end{equation*}
at a specified significance level $\gamma$ and threshold $\epsilon$. We can reject $H_0$
with approximate significance level $\gamma$ and determine MC-EM has converged if
\begin{equation*}
  \Delta \tilde{Q}(\tilde{\params}^{(t,m_t)}, \tilde{\params}^{(t-1)}) + z_{\gamma} \frac{\hat{\eta}}{m_t} < \epsilon.
\end{equation*}

\subsection{Approximate Confidence Intervals}

To quantify uncertainty of the parameter estimates obtained using the MC-EM algorithm,
we adapt the method in \citep{Louis1982} for computing the observed information matrix
in the EM framework. Standard errors, and therefore approximate confidence intervals,
can then be obtained by inverting the observed information matrix. By a result presented
in \citep{Louis1982}, the observed information matrix at a point $\params^{*}$ can be
computed by

\begin{align}
  I(\params^*) = \Big\{ E \Big[- \frac{\partial^2 l(\params)}{\partial \params \partial \params^{T}} \Big| \mathbf{D}, \params^* \Big] & - E \Big[ \Big( \frac{\partial l(\params)}{\partial \params} \frac{\partial l(\params)}{\partial \params^{T}} \Big) \Big| \mathbf{D}, \params^* \Big] \nonumber                                                                     \\
                                                                                                                                       & + E \Big[ \frac{\partial l(\params)}{\partial \params} \Big| \mathbf{D}, \params^* \Big] E \Big[ \frac{\partial l(\params)}{\partial \params^{T}} \Big| \mathbf{D}, \params^* \Big] \Big\} \Big|_{\params = \params^*} \label{eq:I}
\end{align}

where the dependence of $\D$, $\U$ in the log-likelihood has been suppressed for
notational convenience. Let $\hat{\params}$ denote the final MC-EM parameter estimate
and suppose $\{\exposuretrue^j$, $\causal_i^j$, $\clust_i^j: j = 1,\dots, M\}$ are now
importance samples from $P(\exposuretrue, \causal_i, \clust_i | \D, \hat{\params})$ with
(normalized) weights $\{w_i^j\}$. The first and third expectations in \cref{eq:I} can be
approximated by
\begin{align*}
  E \Big[- \frac{\partial^2 l(\params)}{\partial \params \partial \params^{T}} \Big| \mathbf{D}, \params \Big] & \approx \sum_{i=1}^\p \sum_{j=1}^M w_i^j \frac{\partial^2 l_i^j(\params)}{\partial \params \partial \params^T}, \\
  \text{and}\quad  E \Big[ \frac{\partial l(\params)}{\partial \params} \Big| \mathbf{D}, \params \Big]        & \approx \sum_{i=1}^\p \sum_{j=1}^M w_i^j \frac{\partial l_i^j(\params)}{\partial \params}.
\end{align*}

The second expectation in \cref{eq:I} can be approximated by
\begin{align*}
  E \Big[ \Big( \frac{\partial l(\params)}{\partial \params} \frac{\partial l(\params)}{\partial \params^{T}} \Big) \Big| \mathbf{D}, \params^* \Big] & = \sum_{i=1}^\p E \Big[ \Big( \frac{\partial}{\partial \params} l_i(\params) \Big) \Big( \frac{\partial}{\partial \params} l_i(\params) \Big)^T \Big] + 2 \sum_{i < n} E\Big[ \frac{\partial}{\partial \params} l_i(\params) \Big] E\Big[ \frac{\partial}{\partial \params} l_n(\params) \Big]^T                                 \\
                                                                                                                                                      & \approx \sum_{i=1}^\p \sum_{j=1}^M w_i^j \frac{\partial l_i^j(\params)}{\partial \params} \frac{\partial l_i^j(\params)}{\partial \params^T} + 2 \sum_{i < n} \Big[ \sum_{j=1}^M w_i^j \frac{\partial l_i^j(\params)}{\partial \params} \Big] \Big[ \sum_{j=1}^M w_i^j \frac{\partial l_n^j(\params)}{\partial \params^T} \Big].
\end{align*}

We can estimate the standard error of the parameters by inverting the approximated
observed information matrix and taking square root of the diagonal elements. Then,
approximate confidence intervals can be constructed using the asymptotic normality of
maximum likelihood estimates.

\subsection{Probabilistic inference of variant-specific causal
  effects} \label{sec:post-sampling}

To gain a better picture of our knowledge of each individual SNP, we can sample from
$P(\causal_i, \clust_i, \exposuretrue | \exposuredata, \outcomedata, \hat{\params})$---the
posterior distribution of variant-specific latent variables given an MC-EM estimate of
$\params$---by using the sampling/importance resampling (SIR) algorithm \citep{Li2004}.
For example, $P(\clust_i = k | \exposuredata, \outcomedata, \hat{\params})$---the
cluster membership probability of the $i$th SNP---quantifies how certain we are that the
$i$th SNP belongs to a certain cluster. More specifically, suppose we have $M$ samples
$\{\beta_i^j, \clust_i^j, \exposuretrue^j: j = 1,\dots,M \}$ from our importance
distribution for the $i$th SNP. Then we can obtain samples of $\causal_i$ from
$P(\causal_i | \exposuredata, \outcomedata, \hat{\params})$ by sampling---with
replacement---from $\{\causal_i^j\}$ with probabilities proportional to the importance
weights given in \cref{eq:impt-weights}. Samples from
$P(\clust_i | \exposuredata, \outcomedata, \hat{\params})$ and
$P(\exposuretrue | \exposuredata, \outcomedata, \hat{\params})$ can be obtained in a
similar fashion. In particular, these samples can be used to construct credible
intervals of $\causal_i$ and compute cluster membership probabilities---see
\cref{sec:real-data} for examples.

\subsection{Model Selection} \label{sec:model-selection}

To select the number of clusters $K$, we use a modified Bayesian Information criterion
(BIC) for latent variable models estimated using the EM algorithm adopted from
\citet{Ibrahim2008}. For \name, the standard BIC is typically defined as
\begin{equation*}
  \text{BIC} = -2 \log P(\D | \hat{\params}) + (3K + 2p) \log(p)
\end{equation*}

where $\params$ is the MLE of $\params$ and $3K + 2p$ is the dimension of our model. We
replace the marginal density $P(\D | \hat{\params})$ with the readily available IS
estimate of the Q-function at the final MC-EM iteration from \cref{eq:tilde-Q}.

\section{Simulation study} \label{sec:sim}

To verify the efficacy of our statistical inference procedures, we perform two
simulation studies. The goal of the first simulation study is to demonstrate that (1)
the MC-EM algorithm gives parameter estimates that are close to the ground truth and (2)
the approximate confidence intervals we derive have desirable coverage probabilities.
The goal of our second simulation study is to evaluate the accuracy of the modified BIC
for selecting the number of clusters $K$.

\subsection{Parameter Estimation \& Confidence Intervals}
\label{sec:par-est-con-int}

\begin{figure}[ht]
  \begin{subfigure}{0.48\linewidth}
    \includegraphics[width=\linewidth]{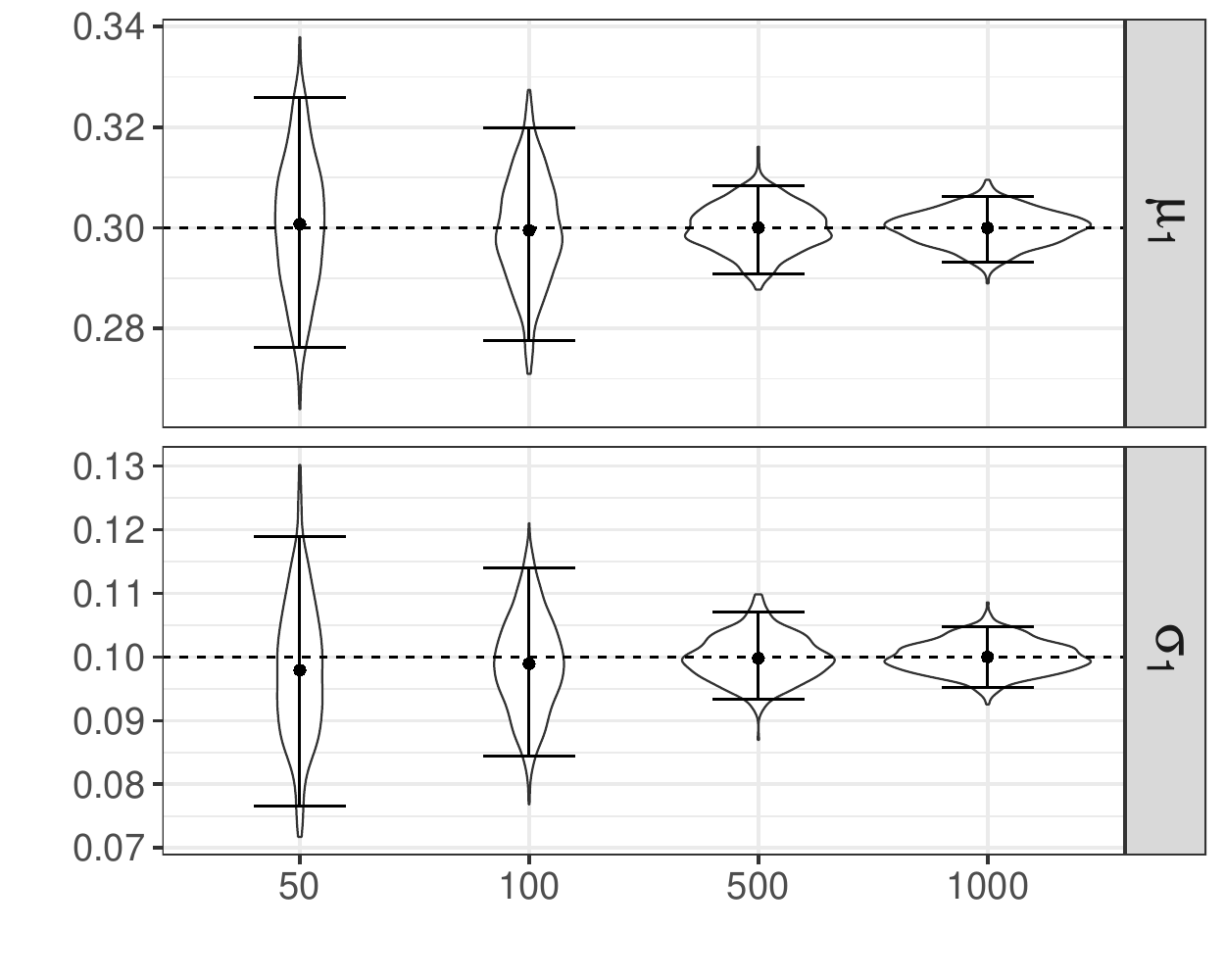}
  \end{subfigure}
  \begin{subfigure}{0.48\linewidth}
    \includegraphics[width=\linewidth]{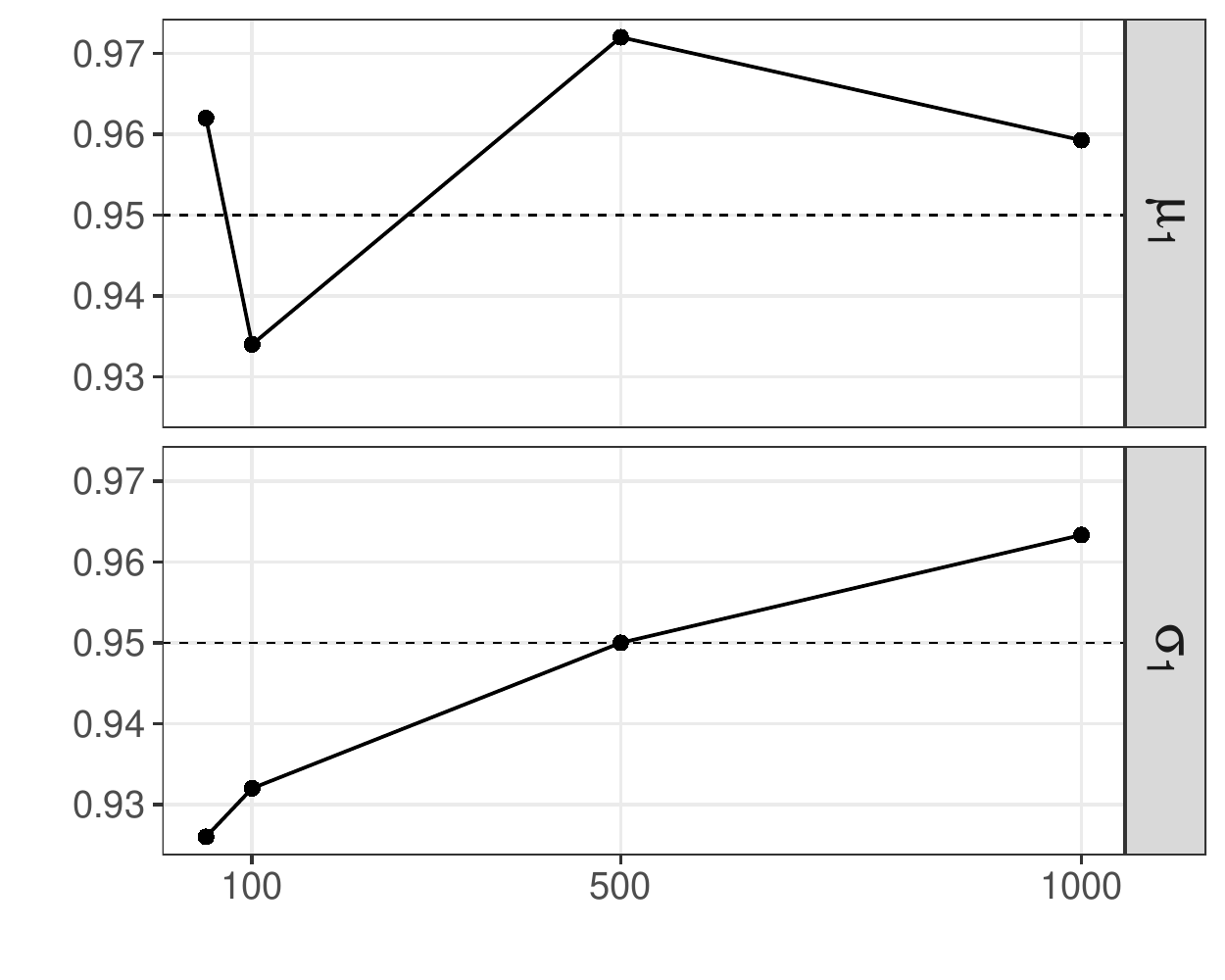}
  \end{subfigure}
  \caption{Simulation study results for $K = 1$ ($\mu_{1} = 0.3$, $\sigma_{1} = 0.1$)
    with 500 replications. \textbf{Left}: Violin plots, 2.5\% and 97.5\% quantiles (solid horizontal line), mean
    (solid point) of parameter estimates as a function of the
    sample size $p$. \textbf{Right}: 95\% coverage probabilities as a function of $p$
    for each parameter.}
  \label{fig:sim-results-K1}
\end{figure}

\begin{figure}[ht]
  \begin{subfigure}{0.48\linewidth}
    \includegraphics[width=\linewidth]{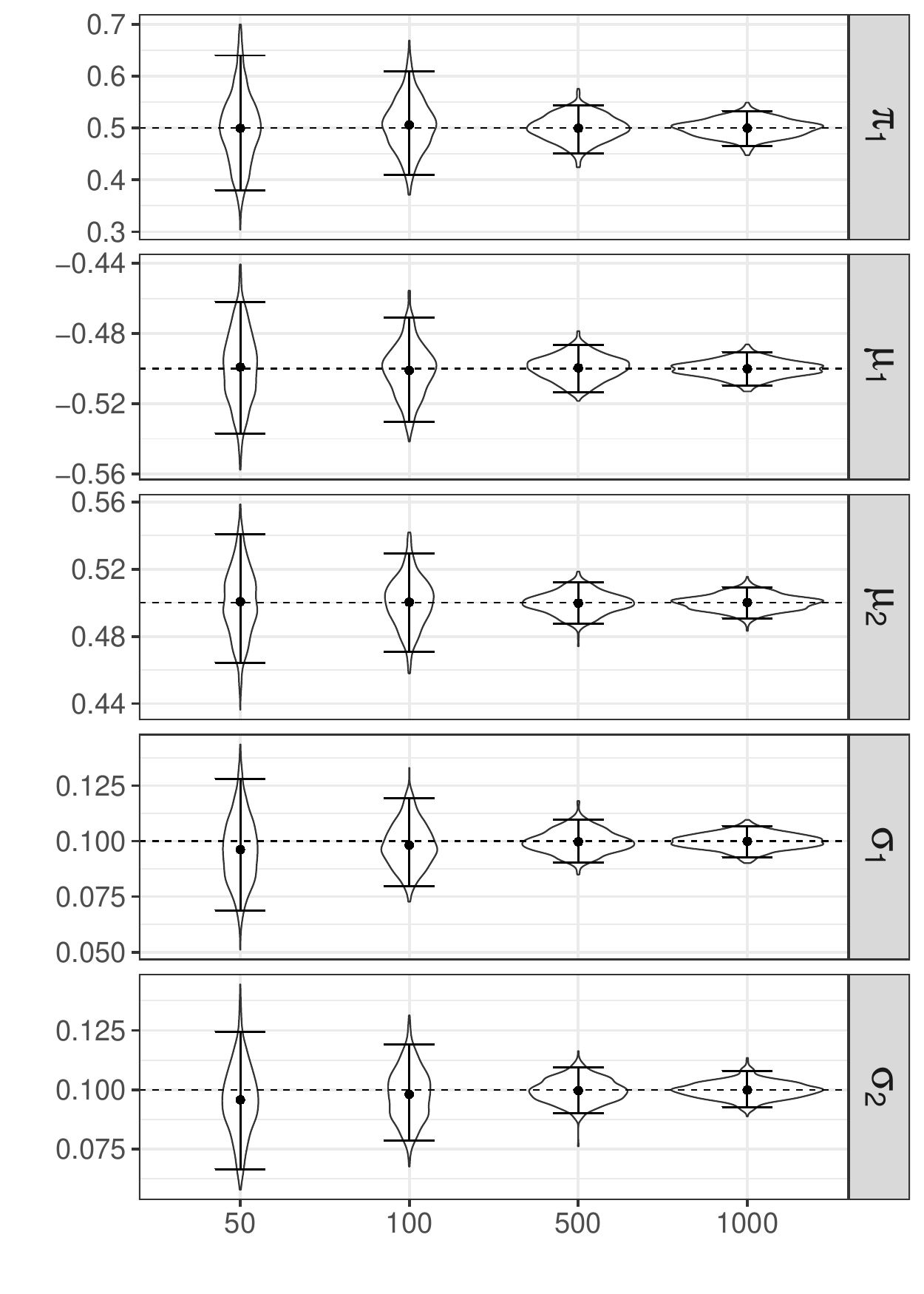}
  \end{subfigure}
  \begin{subfigure}{0.48\linewidth}
    \includegraphics[width=\linewidth]{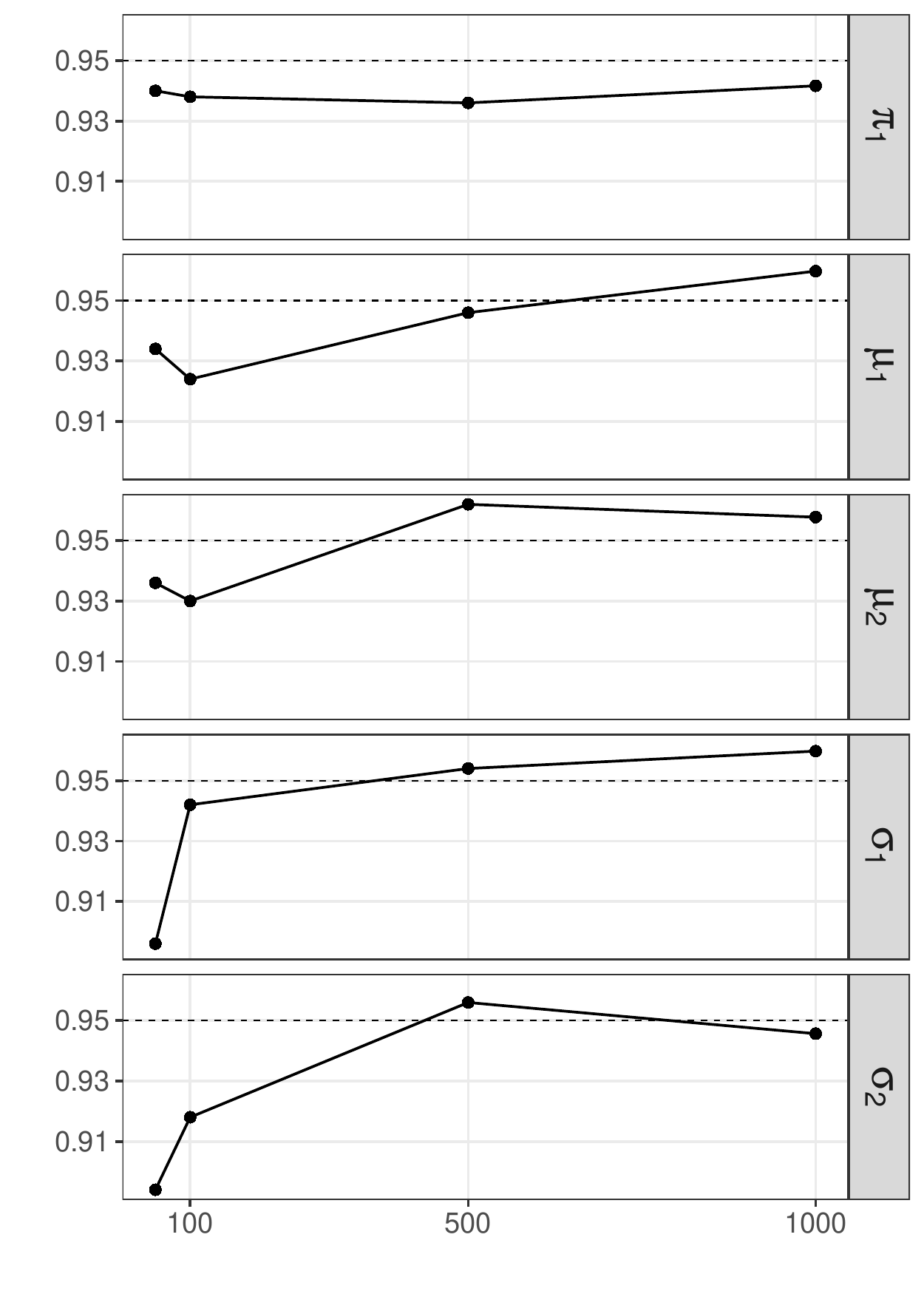}
  \end{subfigure}
  \caption{Simulation study results for $K = 2$ ($\pi_{1} = 0.5$, $\mu_{1} = -0.5$,
    $\mu_{2} = 0.5$, $\sigma_{1} = \sigma_{2} = 0.1$)
    with 500 replications. \textbf{Left}: Violin plots, 2.5\% and 97.5\% quantiles (solid horizontal line), mean
    (solid point) of parameter estimates as a function of the
    sample size $p$. \textbf{Right}: 95\% coverage probabilities as a function of $p$
    for each parameter.}
  \label{fig:sim-results-K2}
\end{figure}

\begin{figure}[ht]
  \begin{subfigure}{0.48\linewidth}
    \includegraphics[width=\linewidth]{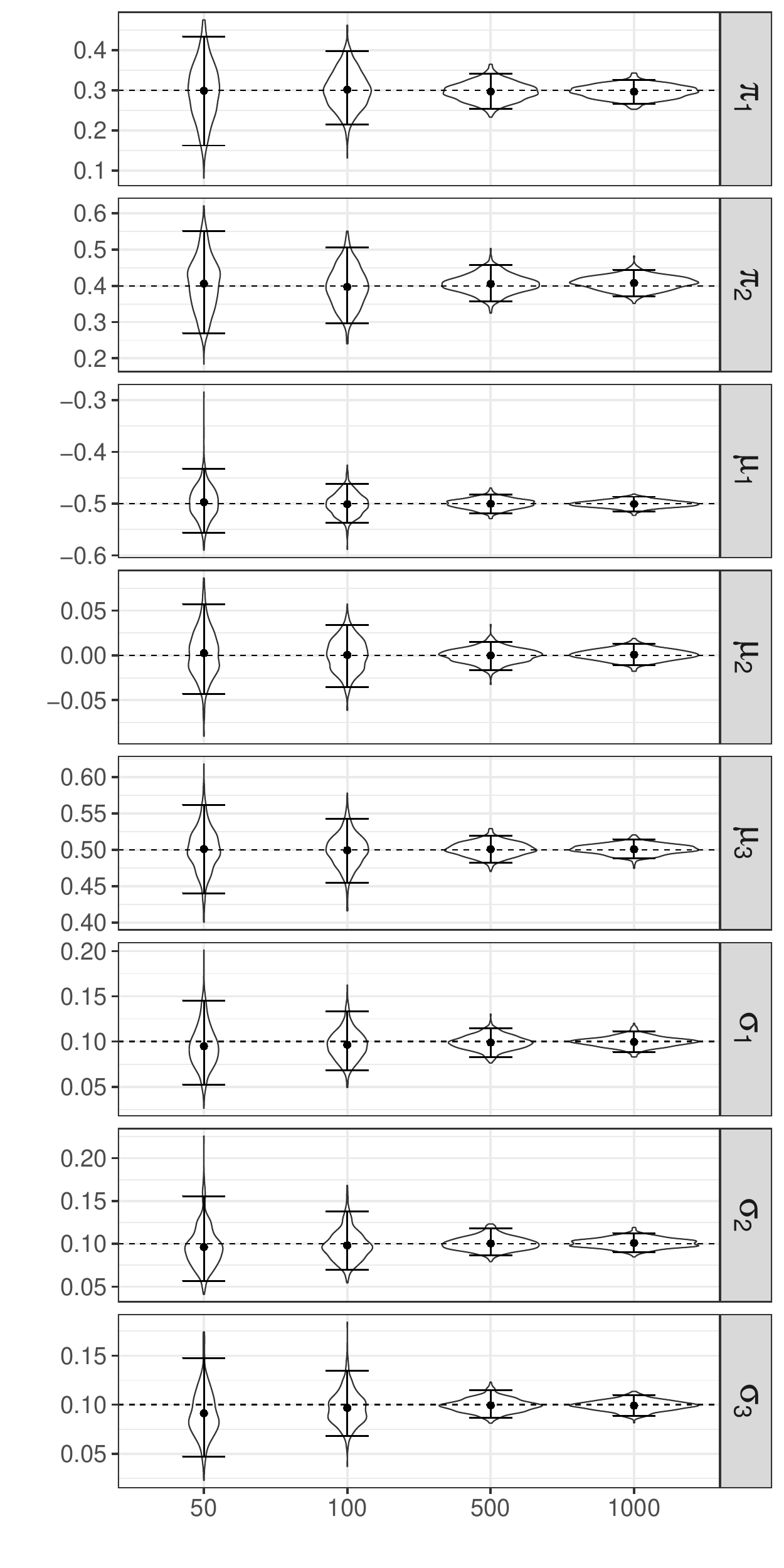}
  \end{subfigure}
  \begin{subfigure}{0.48\linewidth}
    \includegraphics[width=\linewidth]{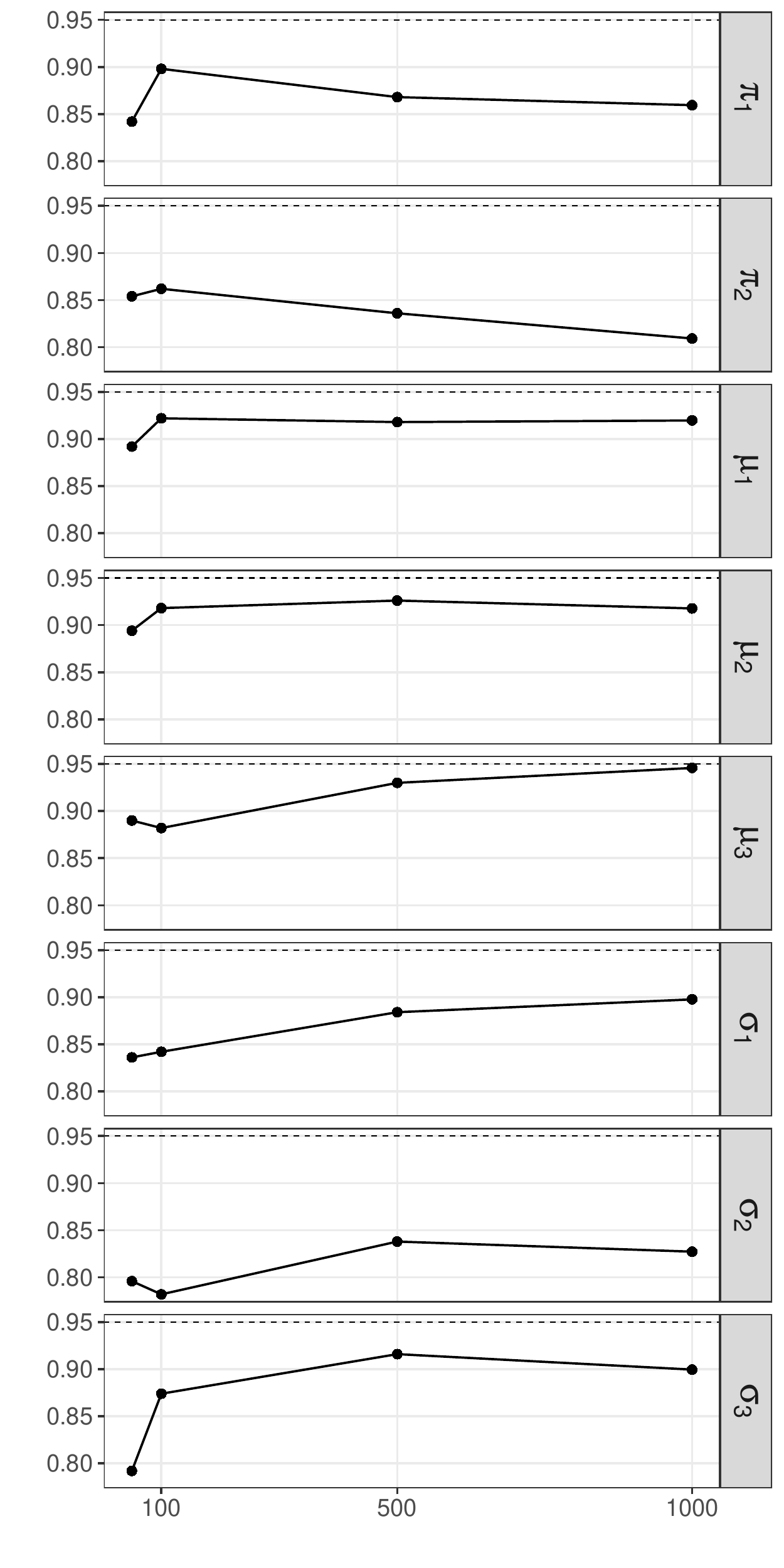}
  \end{subfigure}
  \caption{Simulation study results for $K = 3$ ($\pi_{1} = \pi_{3} = 0.3$, $\mu_{1} = -0.5$,
    $\mu_{2} = 0$, $\mu_{3} = 0.5$, $\sigma_{1} = \sigma_{2} = \sigma_{3} = 0.1$)
    with 500 replications. \textbf{Left}: Violin plots, 2.5\% and 97.5\% quantiles (solid horizontal line), mean
    (solid point) of parameter estimates as a function of the
    sample size $p$. \textbf{Right}: 95\% coverage probabilities as a function of $p$
    for each parameter.}
  \label{fig:sim-results-K3}
\end{figure}

In our first simulation study, we generate simulated data from \name{} under various
parameter settings that mimic GWAS summary data used in practice. In each setting, we
generate measurement errors as $\sigma_{X_i}^2, \sigma_{Y_i}^2 \sim \text{Inv.
  Gamma}(9, .0002)$. This is a reasonable choice as modern GWAS are conducted with large
sample sizes which result in low measurement errors. Moreover, we set the instrument
strength parameter to be $\lambda_x = 10 / \sqrt{p}$ to keep the norm constant across
$p$. We vary the number of genetic variants to be $p = 50, 100, 500, 1000$ and the number of clusters to be $K_{\text{true}} = 1,2,3$. In
MR, we do not expect the number of clusters to be greater than 3 and the number of
filtered genetic variants to be large. For each parameter setting, we ran the MC-EM
algorithm with $K = K_{\text{true}}$ and computed the approximated confidence intervals
with 500 simulated data-sets to obtain parameter estimates and 95\% coverage probabilities.
As the EM algorithm is sensitive to initial value specification,
for each repetition, we ran the MC-EM algorithm 10 times with different random initial
values and report the results from the run with the largest complete-data log
likelihood. We perform a sensitivity analysis for the initial values in the MC-EM
algorithm in \cref{sec:sens-analysis}. Simulation results are
presented in \cref{fig:sim-results-K1,fig:sim-results-K2,fig:sim-results-K3}. Note that
as $K$ increases, we chose mixture means to be closer in value, making the estimation task more challenging.

In most scenarios, the parameter estimates obtained from the 500 replications are
centered around the true value with the variance decreasing as a function of $p$.
Furthermore, the coverage probabilities only deviate at most 5\% from the desired 95\% for
$K = 1, 2$. However, the coverage probabilities for certain parameters are much lower
than the desired level for $K = 3$ even as $p$ increases. Since the parameter estimates are still centered
around the true value, this phenomenon is likely due to underestimation of the standard error.

\subsection{Model Selection with modified BIC}

In our second simulation study, we simulate data from the \name{} model with $p = 50$ or $250$ and
$\sqrt{p} \lambda_x = 1$ or $5$. In each setting, we set the true number of clusters
$K_{\text{true}} = 1$,$2$, or $3$. In each replication, we ran the MC-EM algorithm with
$K = 1, 2, 3$ and chose the $K$ that yields the lowest modified BIC value. Results for
this simulation study are shown in \cref{tab:bic-accuracy}.

\def\arraystretch{1}
\begin{table}[]
  \centering
  \caption{\textit{Results from second simulation study}. For each setting, we simulated
    $N_{\text{rep}} = 500$ data-sets. For $K_{\text{true}} = 1$, we set $\mu_1 = 0.5$
    and $\sigma_1 = 0.1$. For $K_{\text{true}} = 2$, we set $\pi_1 = 0.5$,
    $\mu = (-0.5, 0.5)$, and $\sigma_1 = \sigma_2 = 0.1$. For $K_{\text{true}} = 3$, we
    set $\pi_1 = \pi_2 = 1/3$, $\mu = (-0.5, 0, 0.5)$, and
    $\sigma_1 = \sigma_2 = \sigma_3 = 0.05$. The last 3 columns report the proportion of
    replications for each setting where the modified BIC chose the corresponding $K_{\text{BIC}}$.}
  \begin{tabular}{c c c c c c}
    &                      &                   & \multicolumn{3}{c}{Proportion}               \\
    \cline{4-6}
    $p$                  & $\sqrt{p} \lambda_x$ & $K_{\text{true}}$ & $K_{\text{BIC}} = 1$                        & $K_{\text{BIC}}
                                                                                                       =
                                                                                                       2$
                                                & $K_{\text{BIC}}=3$                                                                                   \\
    \hline
    \multirow{4}{*}{50}  & \multirow{2}{*}{1}   &
                                                  {1}
                                               & \textbf{.84}                  & .06               & .1                                            \\
    \cline{3-6}
    &                      & {2}               & 0                              & \textbf{.83} & .17 \\
    \cline{3-6}          &                      & {3}               & 0                              & .03  & \textbf{.97} \\
    \cline{2-6}
    & \multirow{2}{*}{5}   & {1}               & \textbf{.98}                            & .01 & .01    \\
    \cline{3-6}
    &                      & {2}               & 0                              & \textbf{.95} & .05 \\
    \cline{3-6}          &                      & {3}               & 0                              & .01 & \textbf{.99} \\

    \cline{1-6}

    \multirow{4}{*}{250} & \multirow{2}{*}{1}   &
                                                  {1}
                                               & \textbf{.91}                  & .05              & .04                                            \\
    \cline{3-6}
    &                      & {2}               & 0                              & \textbf{.91} & .09 \\
    \cline{3-6}          &                      & {3}               & .04                              & .03 & \textbf{.93} \\
    \cline{2-6}
    & \multirow{2}{*}{5}   & {1}               & \textbf{1}                              & 0    & 0    \\
    \cline{3-6}
    &                      & {2}               & 0                              & \textbf{1} & 0 \\
    \cline{3-6}          &                      & {3}               & 0                              & 0    & \textbf{1}    \\
  \end{tabular}
  \label{tab:bic-accuracy}
\end{table}

From \cref{tab:bic-accuracy}, we observe that lower values for $\sqrt{p} \lambda_x$
result in the largest decreases in the accuracy of the modified BIC. For example, when
$p = 50$ and $K_{\text{true}} = 2$, the modified BIC chose the correct $K$ 95\% of the
time when $\sqrt{p} \lambda_x = 5$ but only 83\% of the time when
$\sqrt{p} \lambda_x = 1$. We also notice that when $\sqrt{p} \lambda_x = 1$, the
accuracy of the modified BIC decreases with $p$. A practical consequence of this
observation is that the modified BIC is more likely to choose the correct $K$ with few
strong instruments than with more weak instruments.


Furthermore, we simulated addditional data from the \name{} model with
$K_{\text{true}}=1$, $\sqrt{p} \lambda_{x} = 1, 5$ and $\sigma = 1, 5, 10$ to assess how
well the modified BIC is able to correctly identify the true number of clusters as
cluster variance increases. The results for these simulations are given in
\cref{tab:bic-accuracy-largevar}.

To our surprise, the modified BIC criterion is more accurate as the true cluster
variance increases. From \cref{tab:bic-accuracy-largevar}, we can see
that for $K_{\text{true}} = 1$, $p = 50$, and $\sqrt{p} \lambda_{x} = 1$, the modified
BIC criterion chose the correct number of clusters only 84\% of the time when
$\sigma_{1} = 0.1$, whereas it chose correctly 98.8\% of the time when $\sigma_{1} = 1$.
However, the estimates of $\mu_{1}$ across replications where the BIC criterion chooses
$K = 1$ have a mean further away from the true value $0.5$ and a larger standard deviation as $\sigma_{1}$ increases.

\begin{table}[]
  \centering
  \caption{Results from simulation study to assess performance of BIC criterion when
    clusters have high variance. For each setting, we simulated $N_{\text{rep}} = 500$
    data sets with $K_{\text{true}} = 1$ and $\mu_{1} = 0.5$. The first three columns
    display the simulation parameters used in each setting. The next three columns report the proportion
    of replications for each setting where the modified BIC chose the corresponding
    $K_{\text{BIC}}$. The last three columns report the mean and standard deviation of
    the estimated cluster mean $\hat{\mu}_{1}$ across replications in each setting.}

  \begin{tabular}{c c c | c c c | c c}
    &                      &                   & \multicolumn{3}{c}{Proportion} &
                                                                                  \multicolumn{2}{c}{$\hat{\mu}_{1}$
                                                                                  when
                                                                                  $K_{\text{BIC}} = 1$}               \\
    $p$                  & $\sqrt{p} \lambda_x$ & $\sigma_{1}$ & $K_{\text{BIC}} = 1$                        & $K_{\text{BIC}}
                                                                                                       =
                                                                                                       2$
                                                & $K_{\text{BIC}}=3$ & Mean & Std.\ Dev.\                                                                                  \\
    \hline
    \multirow{4}{*}{50}  & \multirow{2}{*}{1}   &
                                                  {0.1}
                                               & .840                  & .062
                         & .098 & .500 & .017                                            \\
    \cline{3-8}  &  &
                                                  {1}
                                               & .988                  & .010
                         & .002 & .508 & .132                                            \\
    \cline{3-8}
    &                      & {5}               & .992                              &
                                                                                     .006
                         & .002 & .538 & .654 \\
    \cline{3-8}          &                      & {10}               & .992
                                                                                & .008
                         & 0 & .621 & 1.300 \\
    \cline{2-8}
    & \multirow{2}{*}{5}   & {0.1}               & .975                            &
                                                                                     .014
                         & .011 & .501 & .013    \\
    \cline{3-8} &   & {1}               & .991                            & .009 & 0 &
                                                                                       .509
                                                               & .131    \\
    \cline{3-8}
    &                      & {5}               & .995                              &
                                                                                     .005
                         & 0 & .552 & .665 \\
    \cline{3-8}          &                      & {10}               & .997
                                                                                & .003 &
                                                                                         0
                                                & .619 & 1.330 \\

    \cline{1-8}

    \multirow{4}{*}{250} & \multirow{2}{*}{1}   &
                                                  {0.1}
                                               & .910                  & .053
                         & .037 & .497 & .010                                            \\
    \cline{3-8} &   &
                                                  {1}
                                               & .996                  & 0
                         & .004 & .499 & .069                                           \\
    \cline{3-8}
    &                      & {5}               & 1                              & 0 & 0
                                                & .505 & .336 \\
    \cline{3-8}          &                      & {10}               & 1
                                                                                & 0 & 0
                                                & .566 & .694 \\
    \cline{2-8}
    & \multirow{2}{*}{5}   & {0.1}               & 1                              & 0
                         & 0 & .500 & .007    \\
    \cline{3-8} &   & {1}               & 1                              & 0    & 0 &
                                                                                      .503
                                                               & .066    \\
    \cline{3-8}
    &                      & {5}               & 1                              & 0 & 0
                                                & .517 & .328 \\
    \cline{3-8}          &                      & {10}               & 1
                                                                                & 0    &
                                                                                         0
                                                & .539 & .658    \\
  \end{tabular}
  \label{tab:bic-accuracy-largevar}
\end{table}

\section{Real data applications} \label{sec:real-data}

\subsection{Results for the motivating HDL-CHD example}
\label{sec:results-motiv-hdl}

We now return to the motivating example introduced in \Cref{sec:motivating-example}. The
dataset being used is created from several large-scale GWAS datasets for plasma lipids
(HDL-C, LDL-C, triglycerides) \citep{teslovich2010biological,willer2013discovery},
coronary heart disease \citep{nikpay2015comprehensive}, lipoprotein subfractions
\citep{kettunen2016genome}, and other cardiovascular diseases. We use the three-sample
summary-data MR design described in \citet{Zhao2018} to preprocess and homogenize the
datasets. We first select 151 independent SNPs (distance $\ge$ 10 mega base pairs,
$R^2 \le 0.001$ in a reference panel) that are associated with at least one plasma lipid
trait (defined as the minimum $p$-value with HDL-C, LDL-C, and triglycerides less than
$10^{-4}$). We then obtain the GWAS associations of these SNPS with all the other
cardiometabolic traits. For the purpose of this example, we will focus on 31 SNPs that
showed genome-wide significant associations ($p$-value $\le 5 \times 10^{-8}$) with
HDL-C in the selection GWAS.

\begin{figure}[t]
  \centering \includegraphics[width=.8\textwidth]{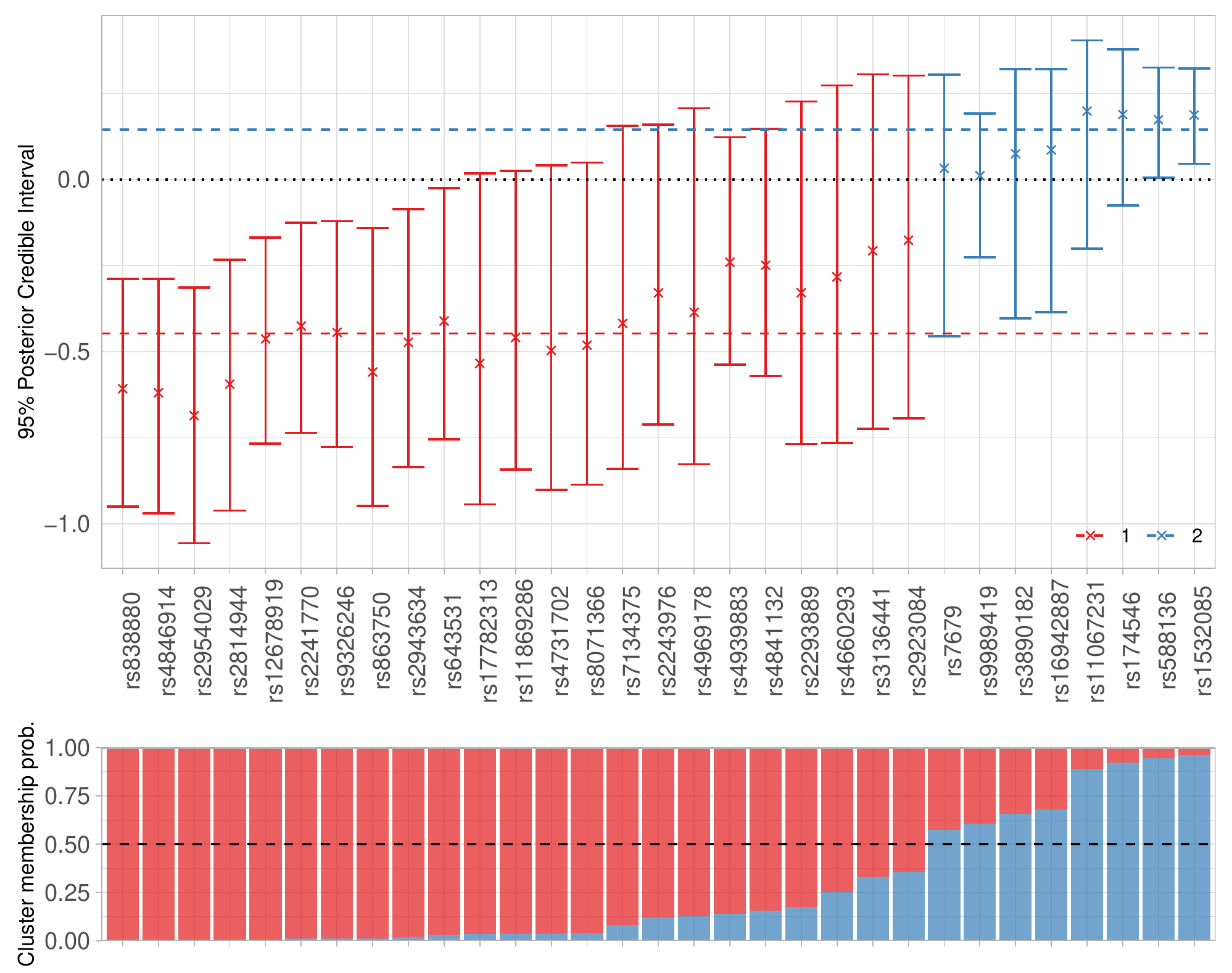}
  \caption{\textit{SNP-specific posterior quantities in HDL-CAD data}. The SNPs are
    ordered by their posterior probability of belonging to cluster 2. \textbf{Top}:
    \textit{95\% posterior credible intervals}. Colored dashed lines are the estimated
    cluster means and x-marks are the posterior medians for each SNP. \textbf{Bottom}:
    \textit{Posterior cluster membership probabilities bar plot}. Vertical axis is
    posterior probability of belonging to cluster 2.}
  \label{plot:hdlcad_bar}
\end{figure}

We apply the Monte Carlo EM algorithm developed in \Cref{sec:inference} to the 31
genetic instruments and their associations with HDL-C in a separate dataset
\citep{kettunen2016genome} and CHD.
The modified BIC from \cref{sec:model-selection} for $K = 1$ and $K = 2$ was -384.72 and
-385.54, respectively, which slightly favors $K = 2$. In other words, the data supports
a model with mechanistic heterogeneity.
The larger cluster ($\hat{\pi}_1 = 0.82$) corresponds to a negative effect
($\hat{\mu}_1 = -0.343$, $\hat{\sigma}_1 = 0.23$) and the smaller cluster
($\hat{\pi}_2 = 0.18$) corresponds to a positive effect ($\hat{\mu}_2 = 0.14$,
$\hat{\sigma}_2=0.09$). The SNPs are classified into the two clusters based on their
posterior probabilities (\Cref{sec:post-sampling}). \Cref{fig:hdlcad-scatter-mixture}
shows the scatterplot of the HDL-CAD data with the MC-EM parameter estimates.
\Cref{plot:hdlcad_bar} shows the posterior estimates of the variant-specific $\beta_j$
along side the cluster membership probabilities.

To validate the mechanistic heterogeneity identified by \name, we generate a heatmap of
the associations (z-scores) of the SNPs with lipoprotein subfraction traits
\citep{kettunen2016genome}. Most of the traits are named after their size (XS = extra
small, S = small, M = medium, L = large, XL = extra large, XXL = double extra large),
their lipoprotein class (HDL, IDL = intermediate-density lipoprotein, LDL, VLDL =
very-low-density lipoprotein), and the measurement (C = total cholesterol, CE =
cholesterol esters, FC = free cholesterol, L = total lipid, P = particle concentration,
PL = phosolipids, TG = triglycerides). Other traits including the mean diameter of
HDL/LDL/VLDL particles (HDL-D/LDL-D/VLDL-D) and the concentration of ApoA1/ApoB (major
protein component of HDL/LDL). To aid visualization, the SNPs are ordered by their
cluster membership probabilities and the lipoprotein subfractions are ordered by their
density and size.

\begin{figure}[t]
  \centering \includegraphics[width = 0.8 \textwidth]{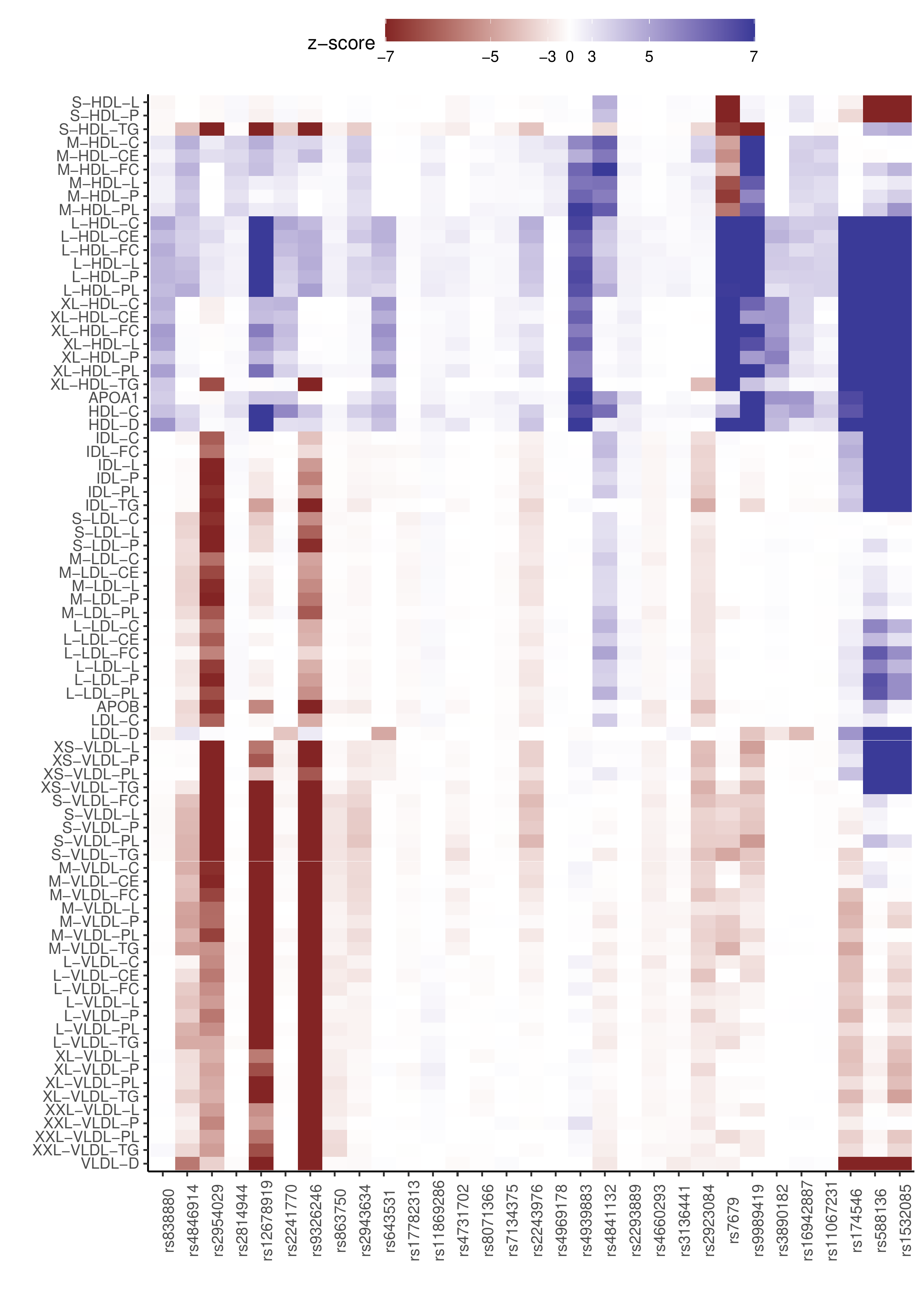}
  \caption{A heatmap showing the z-scores of the SNPs' associations with lipoprotein
    subfraction measurements. The SNPs are ordered by their posterior probability of
    belonging to cluster 2 as in \Cref{plot:hdlcad_bar}. The lipoprotein subfractions
    are ordered by their density and size.}
  \label{plot:hdlcad_heatmap}
\end{figure}

The heatmap in \Cref{plot:hdlcad_heatmap} shows that the SNP clusters found by the
mixture model exhibit different patterns of association with the lipoprotein
subfractions. Several SNPs in the first cluster have strong inverse association with
LDL-C and other LDL/VLDL subfraction traits. Therefore, the negative effect of HDL-C on
CHD suggested by the instruments in the first cluster cluster may indeed be due to their
pleiotropic effect on LDL-C and ApoB-containing lipoproteins (see scenario 1 in
\Cref{fig:multiple-direct-effect}). In contrast, several SNPs in the second cluster
(rs1532085, rs588136, rs174546, rs7679) are inversely associated with the concentration
of small HDL particles, so they may be related to another mechanism that regulates the
size of HDL particles. Although the instruments in this cluster suggest a positive
effect of HDL-C on CHD, this may be explained by heterogeneous effects of cholesterol
contained in different HDL subfractions (see scenario 2 in
\Cref{fig:multiple-direct-effect}). An earlier univariable MR study indeed found that
the concentration of small and medium HDL particles may have a negative effect on CHD,
while the large and extra large HDL particles seem to have no effect \citep{Zhao2019}.
To summarize, the heatmap provides some evidence that the clustering structure
identified by our mixture model indeed corresponds to some distinct underlying
mechanisms.



\subsection{The role of adiposity in type II diabetes}
\label{sec:role-adiposity-type}

\begin{figure}
  \centering
  \begin{subfigure}{.5\textwidth}
    \centering \includegraphics[width=\columnwidth]{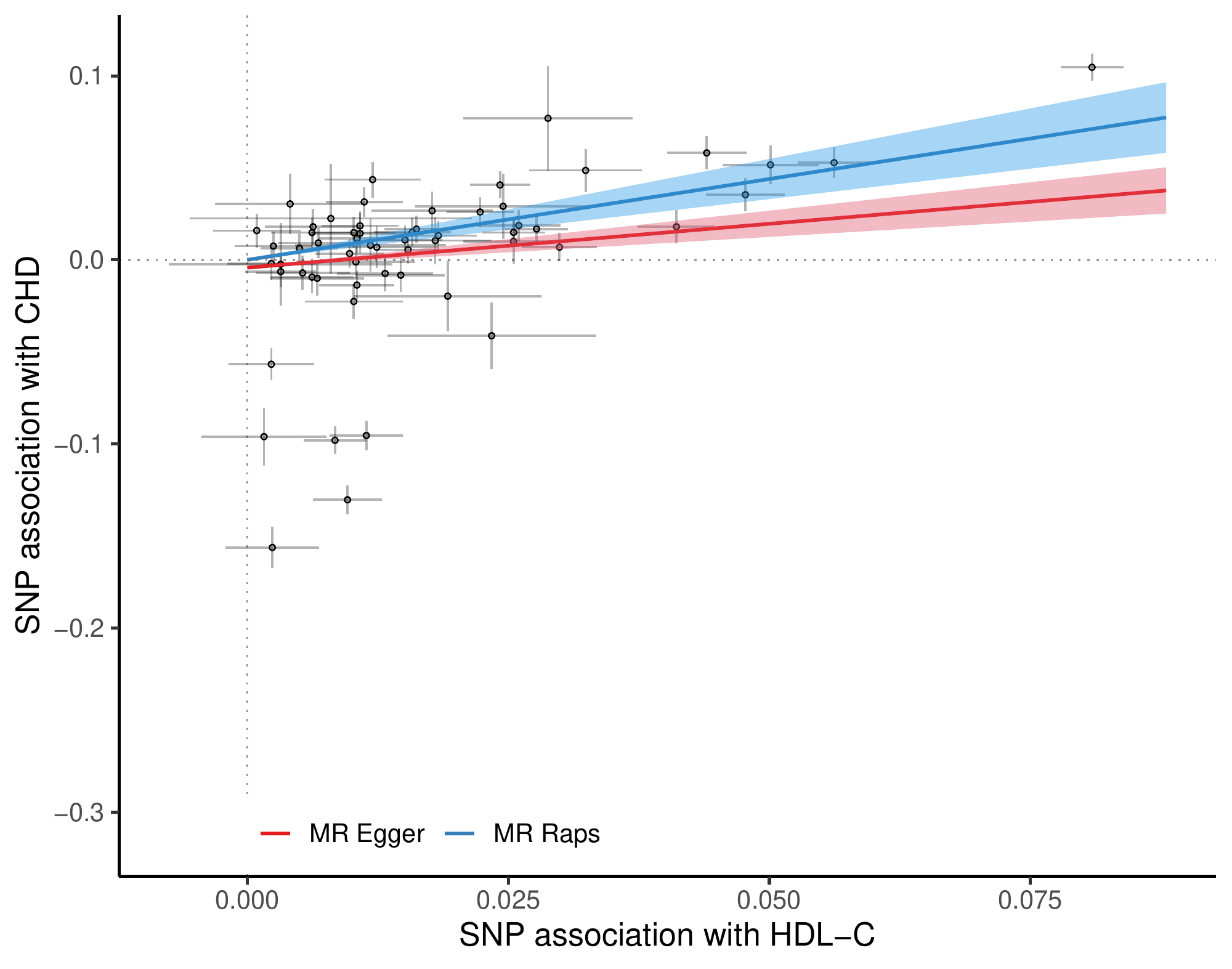}
    \caption{MR-RAPS \& MR-Egger}
    \label{fig:bmi-t2d-scatter-raps}
  \end{subfigure}%
  \begin{subfigure}{.5\textwidth}
    \centering \includegraphics[width=\columnwidth]{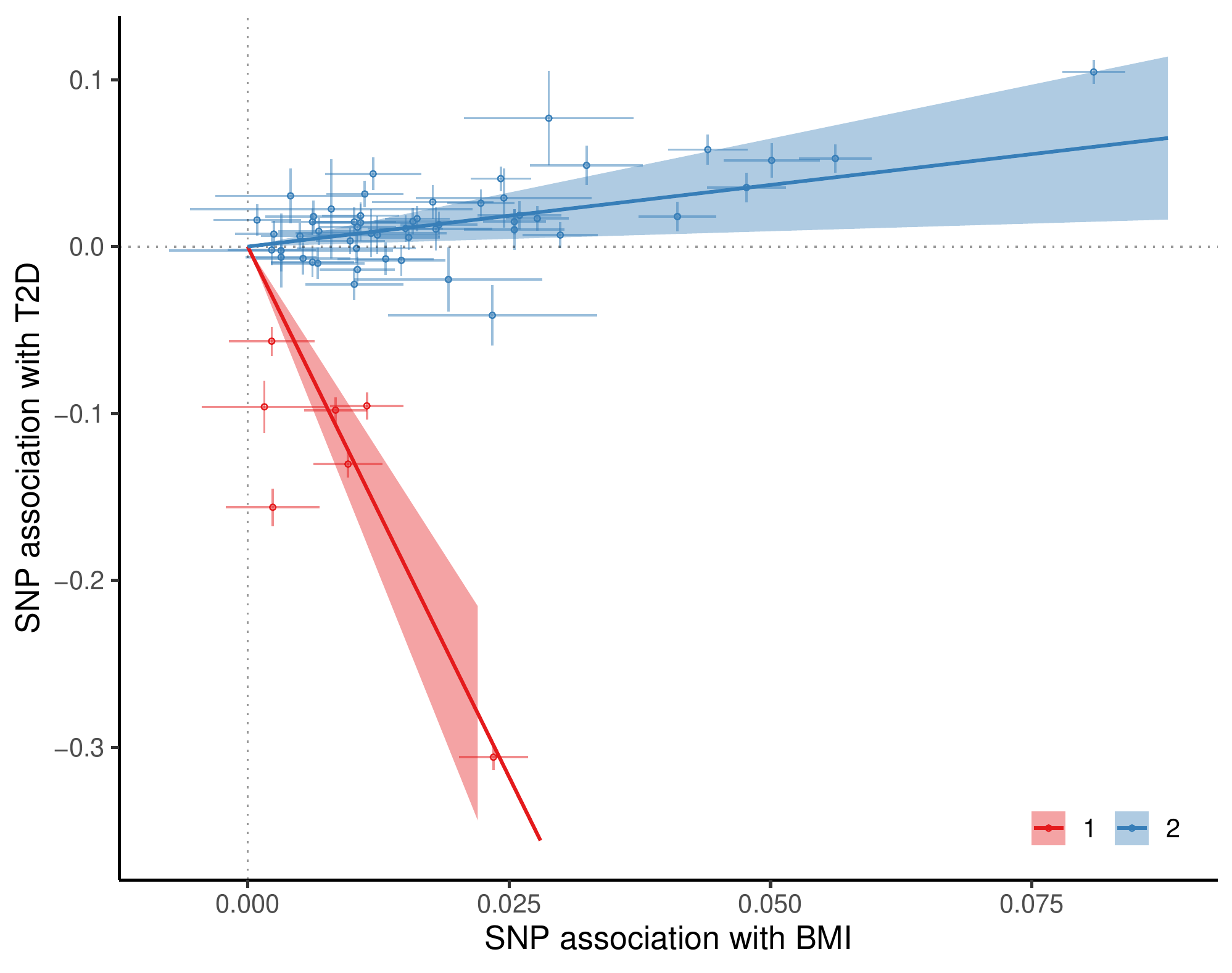}
    \caption{MR-Path}
    \label{fig:bmi-t2d-scatter-mrpath}
  \end{subfigure}
  \caption{\textit{Scatterplot of BMI-T2D data and effect estimates}. \textbf{Left}:
    Line/shaded region represents the causal effect estimate $\pm$ one standard error
    from MR-RAPS and MR-Egger.
    \textbf{Right}: Lines/shaded regions represent heterogeneous causal effect estimates
    $\pm$ one standard deviation from MR-Path.}
  \label{fig:bmi_t2d_scatter}
\end{figure}

We now turn to a second example to illustrate the utility of \name. In this example, we
are interested in possible mechanistic heterogeneity of the effect of adiposity (as
measured by the body mass index, BMI) on type II diabetes (T2D). Following the same
three-sample summary-data MR design as described in \Cref{sec:results-motiv-hdl}, we
created a dataset of 60 SNPs from two GWAS summary datasets for BMI
\citep{akiyama2017genome,locke2015genetic} and one for T2D \citep{mahajan2018fine}. We
then apply the Monte Carlo EM algorithm developed in \Cref{sec:inference}. The modified
BIC selects $K=2$ clusters of SNPs. The larger cluster ($\hat{\pi}_2 = 0.88$)
corresponds to a positive effect ($\hat{\mu}_1 = 0.77$, $\hat{\sigma}_2 = 0.42$) and the
smaller cluster ($\hat{\pi}_1 = 0.12$) corresponds to a very large negative effect
($\hat{\mu}_1 = -12.4$, $\hat{\sigma}_1 = 1.8$). See \Cref{fig:bmi-t2d-scatter-mrpath} for a
scatterplot of the data with effect estimates from MR-Path.

\begin{figure}[t]
  \centering \includegraphics[scale=0.7]{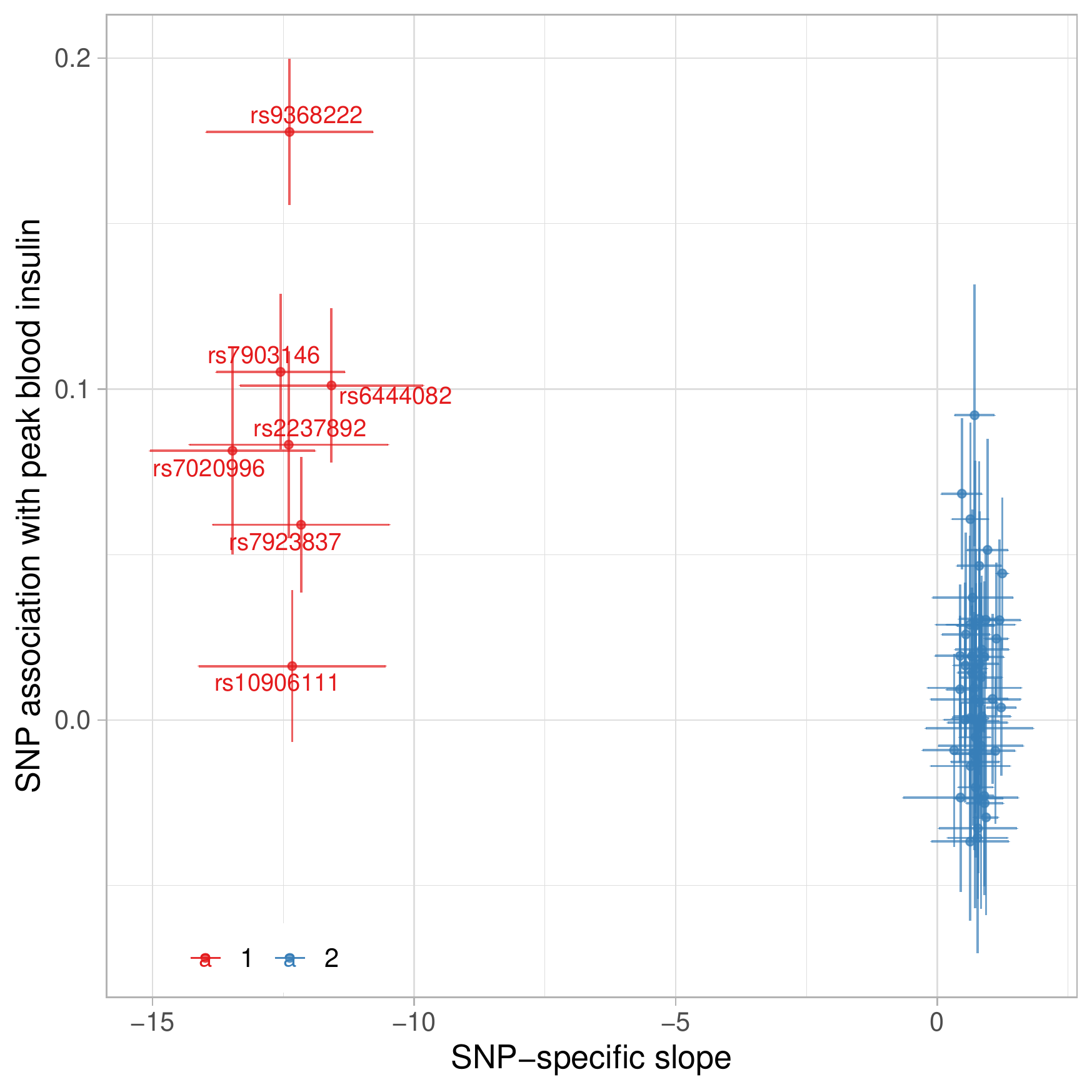}
  \caption{\textit{Relationship of SNP-specific slope and the association with peak
      blood insulin}. The horizontal axis is the posterior median of the SNP-specific
    slope $\beta_i$ with standard error bars. The vertical axis is the SNP association
    with peak blood insulin with standard error bars reported in an independent GWAS.
    The SNPs are colored according to the cluster with the highest posterior cluster
    membership probability $P(\clust_i = k | \exposuredata, \outcomedata)$.}
  \label{plot:bmi_t2d_insulin}
\end{figure}

We did a GWAS catalog \citep{buniello2019nhgri} search for the SNPs belonging to cluster
2 and found that several of them are related to insulin function which tightly regulates
glucose level and plays a crucial role in diabetes. This motivated us to compare the
estimate variant-specific effect $\hat{\beta}_i$ with the SNP association with peak
blood insulin, which is available from an independent GWAS \citep{wood2017genome}
(\Cref{plot:bmi_t2d_insulin}). In fact, six out of the seven SNPs classified into
cluster 2 are strongly associated with peak blood insulin. This shows that the large
negative effect of this cluster is most likely due to horizontal pleiotropy
(\Cref{fig:multiple-direct-effect}) instead of a genuine negative causal effect of
adiposity. The results we obatined here are broadly consistent with other recent genetic
studies that have identified SNPs with opposite effects on adiposity and type II
diabetes and linked the ``favorable adiposity'' genes to insulin function and fat
distribution \citep{ji2019genome}.



\begin{figure}[t]
  \centering \includegraphics[scale=0.7]{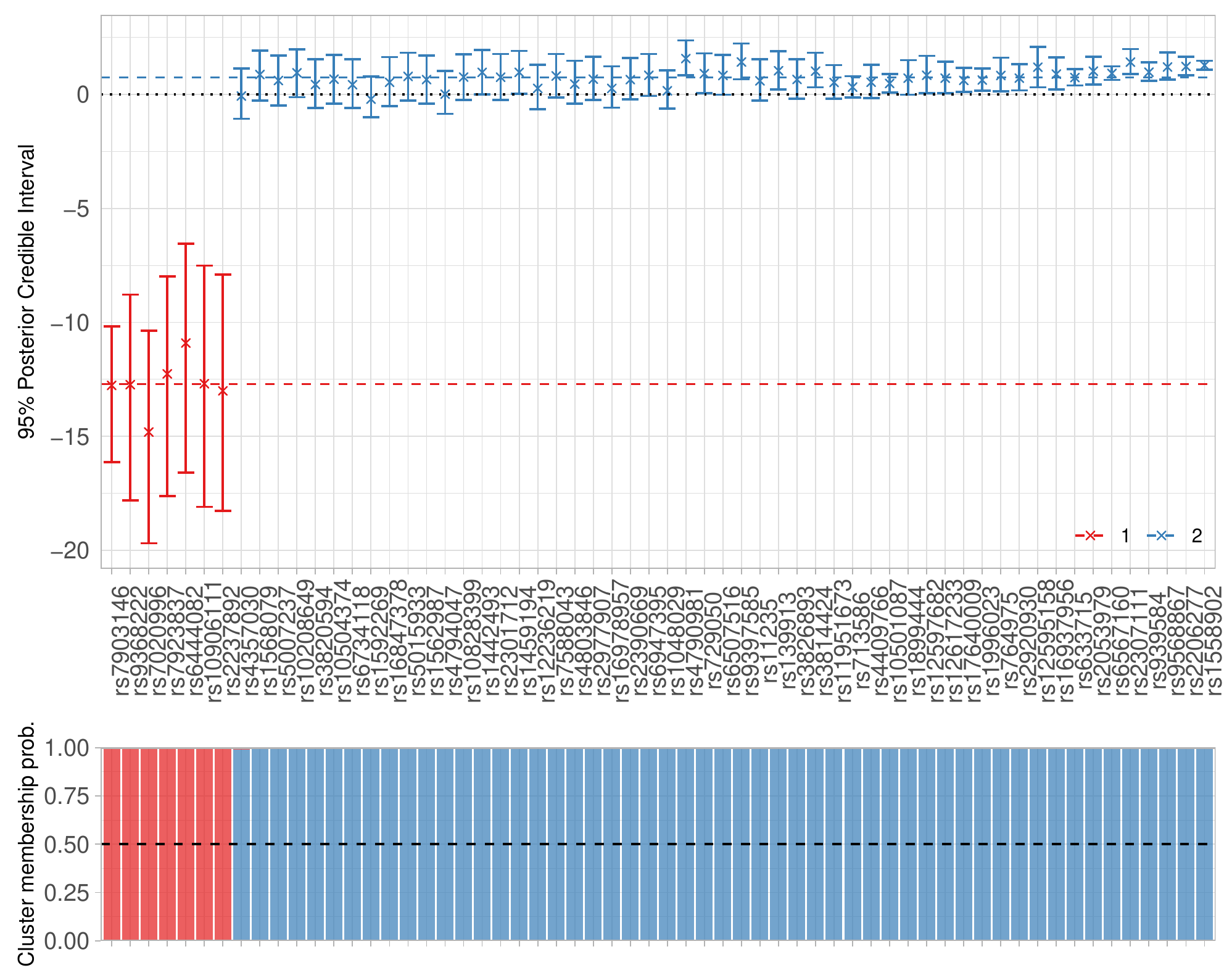}
  \caption{\textit{SNP-specific posterior quantities in BMI-T2D data}. SNPs are ordered
    by their posterior probability of belonging to cluster 2. \textbf{Top}: \textit{95\%
      posterior credible intervals}. Colored dashed lines are the estimated cluster
    means (same as Figure \ref{fig:bmi_t2d_scatter}) and x-marks are the posterior
    medians for each SNP. \textbf{Bottom}: \textit{Posterior cluster membership
      probabilities bar plot}. Vertical axis is posterior probability of belonging to
    cluster 2.}
  \label{plot:bmi_t2d_bar}
\end{figure}

\section{Discussion} \label{sec:discussion}


In this paper, we have formalized the notion of mechanistic heterogeneity in the context
of MR and showed that SNPs on the same biological pathway identify similar causal
effects. Different pathways generally correspond to different causal effects, even if
they are all valid instruments. Motivated by this observation, we introduced MR-Path, an
interpretable mixture model for summary-level GWAS data that can provide valuable
insights on mechanistic heterogeneity.

A conclusion of mechanistic heterogeneity can be used in several ways. If we are in
scenario 1 shown in \cref{fig:multiple-direct-effect}, where heterogeneity is caused by
multiple pathways of horizontal pleiotropy, we can try to identify mediating exposures
for pleiotropic mechanisms and then use a multivariable MR method to effectively remove
the heterogeneity caused by horizontal pleiotropy. For scenario 2 shown in
\cref{fig:multiple-x}, where there are multiple mechanisms for the exposure, genetic
enrichment analysis could be helpful in identifying the upstream pathways. Both cases
rely on post-hoc analyses with external data which is beyond the scope of this paper.
Nevertheless, we illustrated some possibilities in our two real data applications in \cref{sec:real-data}.

MR-Path has several advantages over similar existing methods for MR. First, it relaxes
the effect homogeneity assumption implicit in most existing MR methods so it is able to
identify multiple causal mechanisms. Second, MR-Path does not require substantial domain
knowledge since we use a data-driven approach to select the number of clusters. However,
this means that MR-Path is not able to distinguish between the different mechanisms in
\cref{fig:mechanistic-heterogeneity} without further post-hoc analysis (\cite{Wang2020}). Lastly, MR-Path is based on a full
likelihood and is robust to weak instrument bias since we use an error-in-variables
approach to estimate the variant-specific causal effects.

We showed using numerical simulations that our MC-EM algorithm gives parameter estimates
that are close to the ground truth and the corresponding approximate confidence
intervals have coverage probabilities close to their true values. We also showed that
the modified BIC criterion we used for selecting the number of clusters chose correctly
a majority of the time.

We demonstrated the utility of \name{} in modeling mechanistic heterogeneity in MR
analysis by using it to investigate the causal mechanisms between HDL-C and CHD and
between adioposity and type II diabetes. These examples reinforce the importance of
considering multiple causal mechanisms in MR analysis. Our findings are consistent with
existing genetic studies that use external data. For the HDL-C and CHD data set, \name{}
identifies a cluster with a positive average causal effect which may be associated with
a mechanism that regulates size of HDL particles. In our study of the role of adioposity
on type II diabetes, \name{} finds a cluster with a negative average causal effect that
is likely attributed to horizontal pleiotropy.

Since \name{} is a generative model for multiple causal mechanisms in MR, there are many
potential extensions that can be incorporated in future work. One such extension is to
replace our univariate mixture model with a multivariate model to consider multiple risk
exposures simultaneously. The multivariate version of \name{} can be used to account for
the pleiotropic effects of other lipoproteins in our HDL and CHD example. Another
possible extension is to allow for correlated SNPs by relaxing the independence
assumption in \cref{as:data-model}.

\begin{appendix}


  \section{Derivation of $P(\causal_{i} | \exposuretrue, \outcomedata, \params)$}
  \label{sec:derivation}


  In this section, we will derive \cref{eq:beta-dist} by first deriving
  $P(\causal_{i} | \clust_{i} = k, \exposuretrue, \outcomedata, \params)$ and
  $\tilde{\pi}_{ik} := P(\clust_{i} = k | \exposuretrue, \outcomedata, \params)$, for
  $k = 1,\dots,K$. Then,
  $P(\causal_{i} | \exposuretrue, \outcomedata, \params) = \sum_{k=1}^{K} \tilde{\pi}_{ik} P(\causal_{i} | \clust_{i} = k, \exposuretrue, \outcomedata, \params)$.
  For notational convenience, we will drop the dependence on model parameters $\params$.

  \begin{align}
    P(\causal_{i} | \clust_{i} = k, \exposuretrue, \outcomedata)
    &\propto P(\outcomedata | \exposuretrue, \causal_{i}) P(\causal_{i} | \clust_{i} = k) \nonumber \\
    &\propto \exp\Big\{-\frac{1}{2} \Big[ \sigma_{Y_{i}}^{-2} (\outcomedata - \causal_{i} \exposuretrue)^{2} + \sigma_{k}^{-2} (\causal_{i} - \mu_{k})^{2}  \Big] \Big\} \nonumber \\
    &\propto \exp\Big\{ (\sigma_{Y_{i}}^{-2} \exposuretrue^{2} + \sigma_{k}^{-2}) \causal_{i}^{2} - 2 (\sigma_{Y_{i}}^{-2} \exposuretrue \outcomedata + \sigma_{k}^{-2} \mu_{k}) \causal_{i}  \Big\}. \label{eq:deriv-eq-1}
  \end{align}
  It follows from completing the square that
  $\causal_{i} | \clust_{i} = k, \exposuretrue, \outcomedata \sim N(\tilde{\mu}_{ik}, \tilde{\sigma}_{ik}^{2})$,
  where $\tilde{\mu}_{ik}$ and $\tilde{\sigma}_{ik}^{2}$ are given in
  \cref{eq:beta-dist-params}.

  \begin{align*}
    P(\clust_{i} = k | \exposuretrue, \outcomedata)
    &\propto P(\clust_{i} = k) P(\outcomedata | \clust_{i} = k, \exposuretrue) \\
    &\propto \pi_{k} \int P(\outcomedata | \exposuretrue, \causal_{i}) P(\causal_{i} | \clust_{i} = k) d\causal_{i}. \\
  \end{align*}
  From \cref{eq:deriv-eq-1}, we have that
  $\int P(\outcomedata | \exposuretrue, \causal_{i}) P(\causal_{i} | \clust_{i} = k) d\causal_{i} = \Big[(\sigma_{Y_{i}}^{-2}\exposuretrue^{2} + \sigma_{k}^{-2})\Big]^{1/2}$.
  Then, $P(\clust_{i} = k | \exposuretrue, \outcomedata) = \tilde{\pi}_{ik}$.

  \section{Bounded Importance Sampling Weights}
  \label{sec:IS-weights-bound}

  Following \cref{eq:target-distr}, the importance weights are given by

  \begin{align*}
    w_i^j & = \frac{P(\exposuretrue | \exposuredata, \outcomedata,
            \params)}{P(\exposuretrue | \exposuredata, \params)}                                                                                                                                                                                                \\
          & \propto P(\exposuretrue | \exposuredata, \params)
            P(\outcomedata | \exposuretrue, \params)                                                                                                                                                                                                            \\
          & = \sum_{k=1}^K \frac{\pi_k}{\sqrt{2\pi \exposuretrue^2 \sigma_k^2 + \sigma_{Y_i}^2}} \exp \Big\{-\frac{1}{2(\exposuretrue^2 \sigma_k^2 + \sigma_{Y_i}^2)} (Y_i - \exposuretrue \mu_k)^2  \Big\} \leq \frac{1}{\sqrt{2 \pi \sigma_{Y_i}^2}},
  \end{align*}

  for $i = 1,\dots,p; j=1,\dots,M$. Therefore, the importance weights are bounded and
  have a finite variance.

  \section{Importance sampling estimate of $\eta^2$}
  \label{sec:IS-eta}

  An importance sampling estimate of $\eta^2$ at iteration $t$ in
  \cref{eq:deltaQ-asymp-dist} is given by

  \begin{align*}
    \hat{\eta}^2 = m_t \sum_{i=1}^p \Big\{ \sum_{j=1}^{m_t} w_i^j \Lambda_{ij}^{(t)} \Big\}^2 \Big[ \frac{\sum_{j=1}^{m_t} (w_i^j \Lambda_{ij}^{(t)})^2 }{(\sum_{j=1}^{m_t} w_i^j \Lambda_{ij}^{(t)})^2} - 2 \frac{\sum_{j=1}^{m_t} (w_i^j)^2 \Lambda_{ij}^{(t)}}{\sum_{j=1}^{m_t} w_i^j \Lambda_{ij}^{(t)}} + \sum_{j=1}^{m_t} (w_i^j)^2 \Big],
  \end{align*}

  where $\Lambda_{ij}^{(t)}$ is given in \cref{eq:lambda-ij}.

  \section{Comparison with MR-Clust}
  \label{sec:comparison-with-mr}

  In this section, we will compare MR-Path with a similar method for identifying
  heterogeneity in Mendelian Randomization known as MR-Clust (\cite{Foley2019}). A
  fundamental difference between MR-Path and MR-Clust is that the former assumes an
  error-in-variables regression model for the observed instrument-exposure and
  instrument-outcome associations (\cref{as:data-model}) and models variant-specific
  causal effects as a latent variable that follows a mixture distribution, while the
  latter models the Wald ratio estimates
  $\hat{\theta}_{i} = \outcomedata / \exposuredata$ using a mixture model.
  More specifically, MR-Clust makes the following assumption:
  \begin{equation}
    \label{eq:mr-clust-model}
    \hat{\theta}_{i} | \{\Theta, \hat{\sigma}_{i}^{2}, \clust_{i} = k\} \sim N(\mu_{k}, \hat{\sigma}_{i}^{2}), \ \text{for
    } k = 1,\dots,K,
  \end{equation}
  where $\hat{\sigma}_{j}$ is the standard error of the $j$th ratio estimate and
  $\Theta$ is a vector of cluster means. Furthermore, MR-Clust assumes there are $K + 2$
  clusters of genetic variants, with $K$ substantive clusters, a null cluster, and a
  junk cluster. The null cluster is assumed to have mean $\mu_{0} = 0$. The junk cluster
  follows a generalized t-distribution in order to account for the remaining genetic
  variants that do not belong to any other cluster. Similar to MR-Path, MR-Clust
  determines the number of clusters $K$ using the Bayesian information criterion (BIC).

\begin{figure}[ht]
    \centering
    \begin{subfigure}{0.48\linewidth}
      \centering \includegraphics[width=\columnwidth]{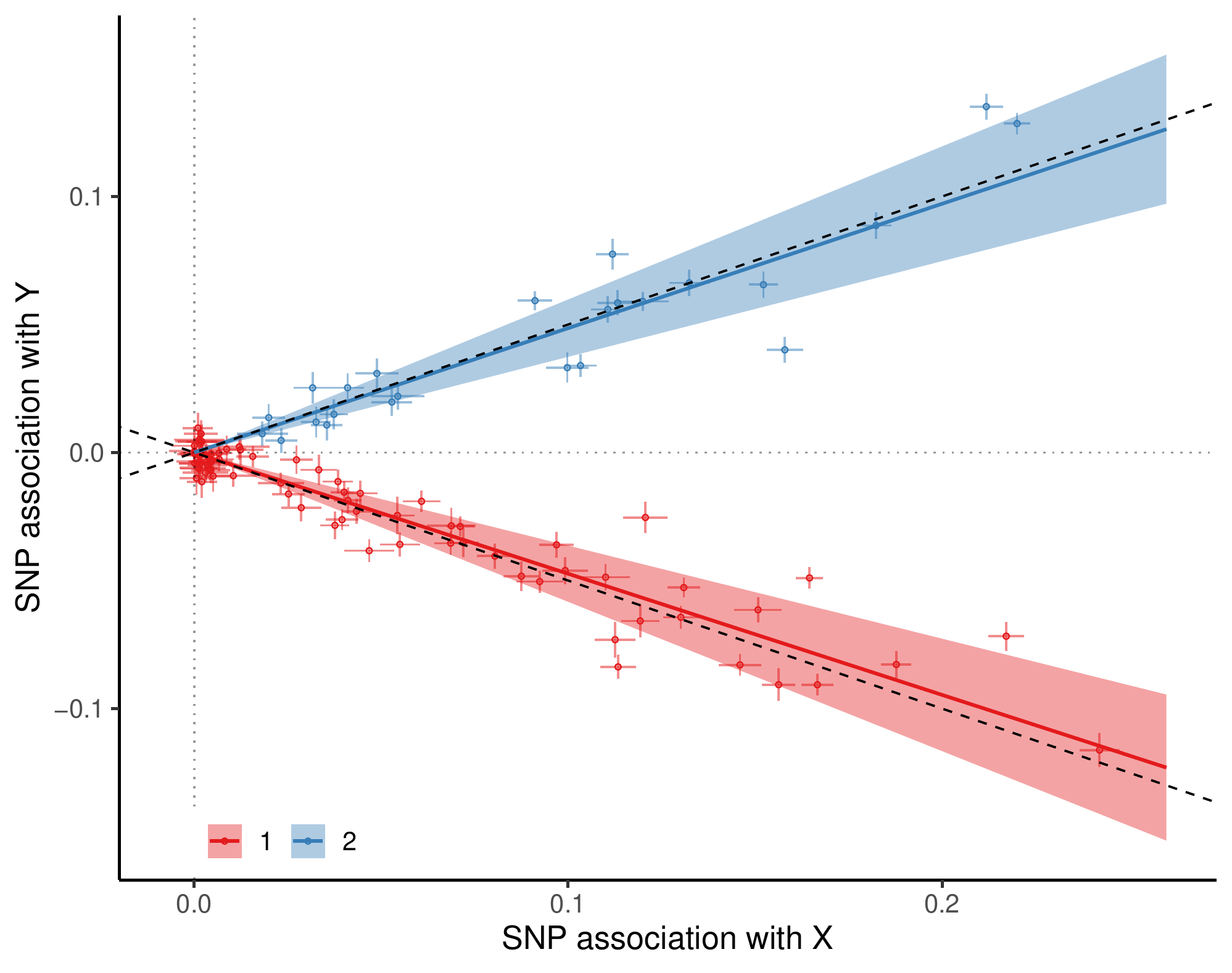}
      \caption{MR-Path}
    \end{subfigure}
    \begin{subfigure}{0.48\linewidth}
      \centering \includegraphics[width=\columnwidth]{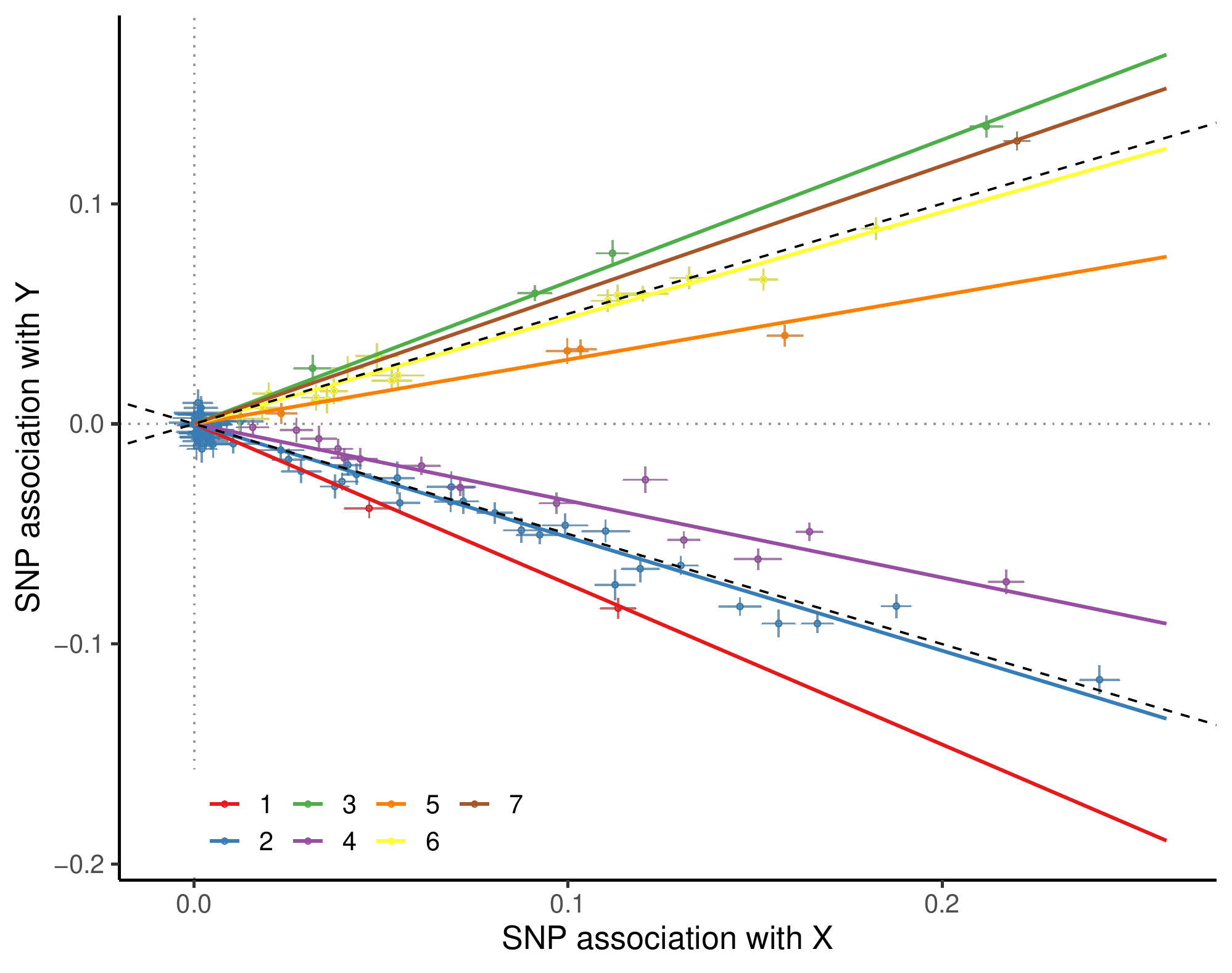}
      \caption{MR-Clust}
    \end{subfigure}
    \caption{Scatterplot of simulated data with many weak instruments and effect
      estimates from MR-Path (left) and MR-Clust (right). The dashed black
      line shows the true cluster means $(-0.5, 0.5)$.}
    \label{fig:weak-instr-bias-scatters}
  \end{figure}

  A downside of directly modeling ratio estimates is that they can be heavily biased for
  weak instruments (\cite{Zhao2018}), increasing the risk of detecting spurious
  clusters. By using an errors-in-variables regression approach, MR-Path is more robust
  to this weak instrument bias. To illustrate this, we simulated data from the model
  described in \cref{sec:mech-het-model} with $p = 100$,
  $\exposuretrue \sim 0.7 N(0, 0.1) + 0.3 N(0, 0.000001)$, $\pi = (0.6, 0.4)$,
  $\mu = (-0.5, 0.5)$, and $\sigma = (0.1, 0.1)$. The estimates from MR-Path and
  MR-Clust are plotted in \cref{fig:weak-instr-bias-scatters}. MR-Path chooses $K = 2$
  (by varying $K$ from 1 to 7 and picking the one with lowest modified BIC), while
  MR-Clust chooses $K = 7$ using a similar model selection procedure. Another advantage
  of MR-Path over MR-Clust is that it constructs confidence intervals for the cluster means.

  However, MR-Clust is more computationally efficient since the parameters in
  \cref{eq:mr-clust-model} can be estimated using an exact EM algorithm. There are
  several ways to close this gap in computational efficiency that we will explore in
  future work. One possibility is to replace the MC-EM algorithm with a variational EM
  algorithm (\cite{blei2017variational}). However, finding a suitable variational approximation to the E-step may be challenging.

\begin{figure}[ht]
    \centering
    \begin{subfigure}{.5\textwidth}
      \centering \includegraphics[width=\columnwidth]{plots/hdlcad_scatter.pdf}
      \caption{MR-Path}
      \label{fig:compare-scatter-mrpath}
    \end{subfigure}%
    \begin{subfigure}{.5\textwidth}
      \centering \includegraphics[width=\columnwidth]{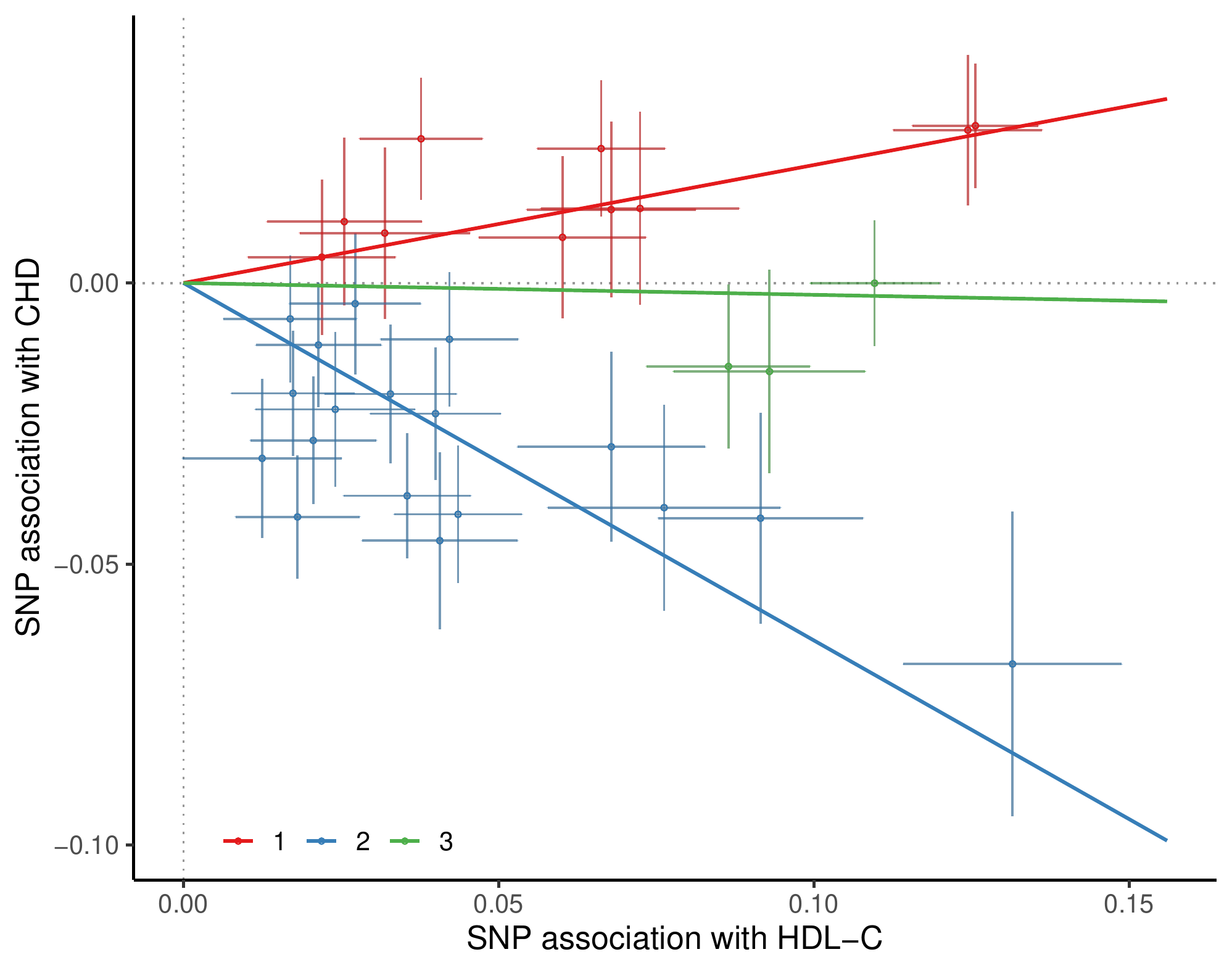}
      \caption{MR-Clust}
      \label{fig:compare-scatter-mrclust}
    \end{subfigure}
    \caption{Scatterplot of HDL-CAD data similar to \cref{fig:hdlcad-scatter} showing
      results from MR-Path (left) vs. MR-Clust (right). The 3rd cluster (green) from
      MR-Clust is the null cluster.}
    \label{fig:hdlcad-scatter-compar}
  \end{figure}

  The results from MR-Path and MR-Clust applied to the motivating HDL-CAD data are
  plotted in \cref{fig:hdlcad-scatter-compar}. In this case, MR-Clust detects two
  substantive clusters with means $0.21$ and $-0.64$ and one null cluster with mean
  $-0.021$. The two substantive clusters are similar to the two clusters detected by
  MR-Path.



\section{Robustness of MR-Path under pleiotropy}
\label{sec:robustn-under-plei}

  To assess the robustness of MR-Path under pleiotropy, we simulate data from the model
  below and conduct a study similar to the one in \cref{sec:par-est-con-int}.
  \begin{align*}
    P(\causal_{i} = \mu_{k}) &= \pi_{k}, \ k = 1,\dots,K, \\
    \exposuretrue &\sim N(0, \lambda_{x}^{2}), \\
    \begin{pmatrix}
      \exposuredata \\ \outcomedata
    \end{pmatrix} \Big| \ \exposuretrue, \causal_{i}, \alpha_{i} &\sim N\Big(
                                                                   \begin{pmatrix}
                                                                     \exposuretrue \\ \alpha_{i} + \beta_{i} \exposuretrue
                                                                   \end{pmatrix},
    \begin{pmatrix}
      \sigma_{X_{i}}^{2} & 0 \\ 0 & \sigma_{Y_{i}}^{2}
    \end{pmatrix} \Big),
  \end{align*}
  where $\alpha_{i}$ represent the direct effect of SNP $i$ on the outcome. Similar to
  \cite{Zhao2018}, we generate $\alpha_{i}$ in three different ways:
  \begin{enumerate}
    \item \textbf{Normal}: $\alpha_{i} \sim N(0, \tau_{0}^{2})$.
    \item \textbf{Laplace}: $\alpha_{i} \sim \tau_{0} \cdot \text{Lap}(1)$, where $\text{Lap}(1)$ is the
          Laplace (double exponential) distribution with rate $1$.
    \item \textbf{Idiosyncratic}: $\alpha_{i}$ is generated according to setup 1 above, except that for 10\% of
          randomly selected SNPs, $\alpha_{i} \sim N(5 \cdot \tau_{0}, \tau_{0}^{2})$.
  \end{enumerate}

\begin{figure}[ht]
    \centering
    \includegraphics[width=\linewidth]{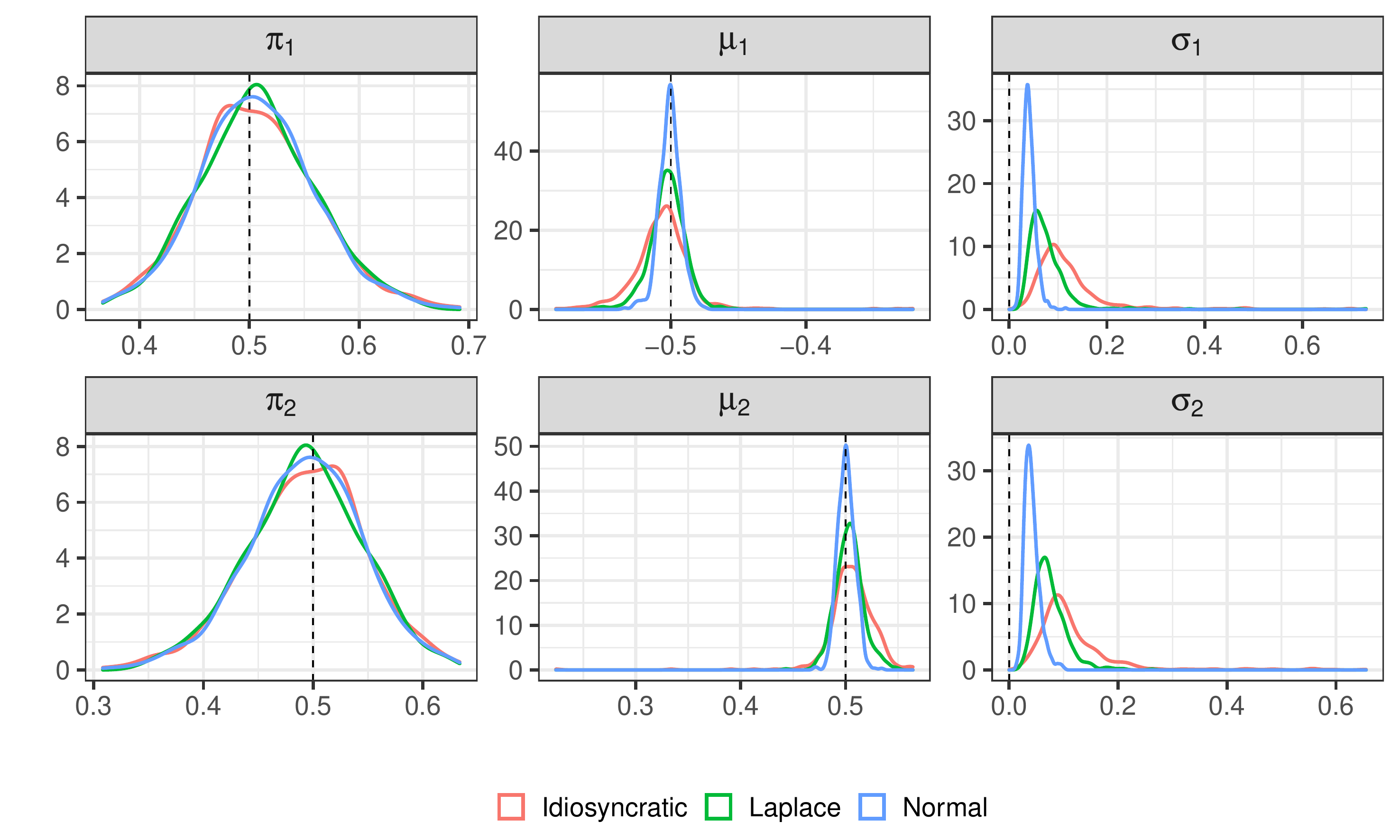}
    \label{fig:sim1-param-est-dens}
    \caption{Density plots for MR-Path estimates of each parameter under different
      scenarios of pleiotropy.}
  \end{figure}

  In each of the scenarios above, we set
  $\tau_{0} = (2 / p) \sum_{j=1}^{p} \sigma_{Y_{i}}$. We generate measurements errors $\sigma_{X_{i}}^{2}$ and $\sigma_{Y_{i}}^{2}$ from the
  same distribution in \cref{sec:par-est-con-int} and set $p = 100$,
  $\lambda_{x} = 10 / \sqrt{p}$. Furthermore, we set $K = 2$, where $\pi_{1} = 0.5$,
  $\mu_{1} = -0.5$, and $\mu_{2} = 0.5$. Density plots for the parameter estimates across 500 replications under
  each scenario above are shown in \cref{fig:sim1-param-est-dens}. For each scenario,
  the estimates of $\pi_{1}$, $\mu_{1}$, and $\mu_{2}$ across replications are centered
  around the true value with increasing variance as we go from normally distributed
  $\alpha_{i}$ to idiosyncratic $\alpha_{i}$. This suggests that our proposed method is
  robust to different types of pleiotropy.

  \section{Computational efficiency of Monte-Carlo EM algorithm}
  \label{sec:comp-efficiency}

  For the HDL-CHD example in \cref{sec:results-motiv-hdl}, the MC-EM algorithm,
  including initial value optimization and model selection, took approximately 6 seconds
  to run on a Dell XPS-15 laptop with an Intel Core i7-8750H processor and 16GB of RAM.
  However, the MC-EM algorithm took roughly 2 minutes to run for the BMI-T2D example in
  \cref{sec:role-adiposity-type}. This is because the algorithm converged much quicker
  for the HDL-CHD example. The BMI-T2D example required an average of 43 iterations and
  80,000 Monte-Carlo samples at termination for each repetition, while the HDL-CHD example
  only required an average of 20 iterations and 5800 Monte-Carlo samples at termination.
  The computational bottleneck of the MC-EM algorithm is its memory usage since it
  requires saving a large matrix of importance samples in the E-step which grows in size
  with each iteration.



  \section{Sensitivity Analysis for Monte-Carlo EM algorithm}
  \label{sec:sens-analysis}

\begin{figure}[ht]
    \centering
    \begin{subfigure}{0.45\linewidth}
      \includegraphics[width=\linewidth]{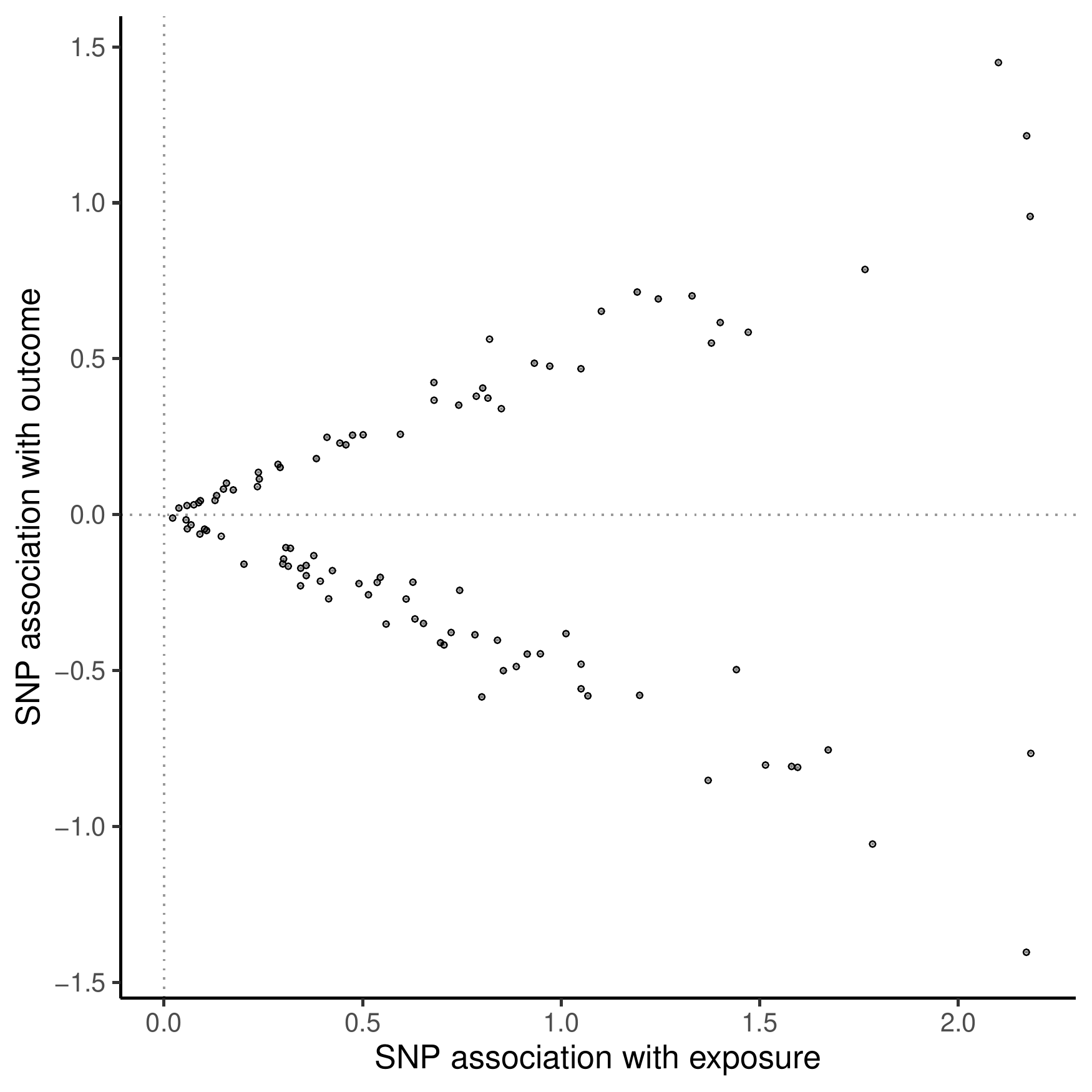}
      \caption{Low overlap}
    \end{subfigure}
    \begin{subfigure}{0.45\linewidth}
      \includegraphics[width=\linewidth]{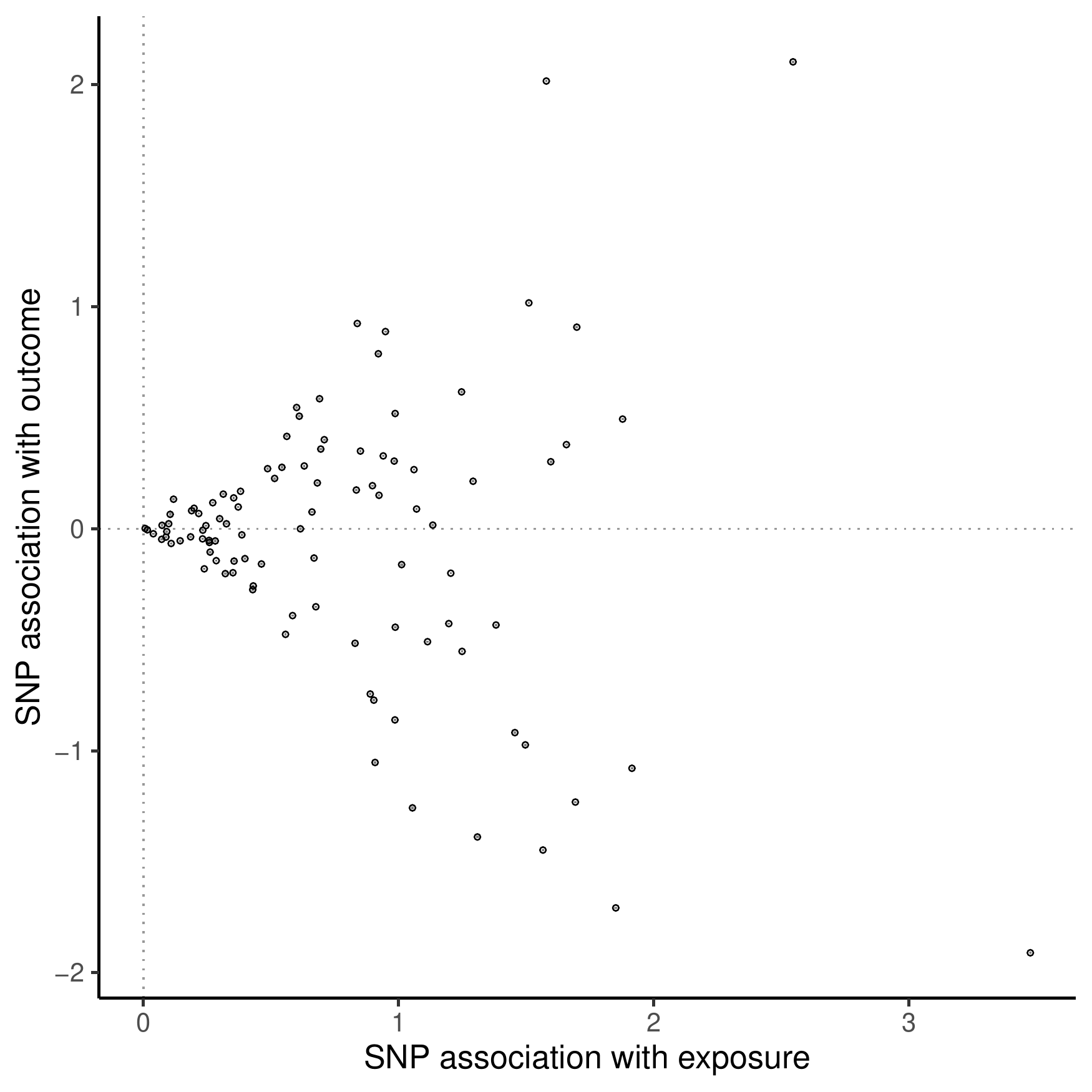}
      \caption{High overlap}
    \end{subfigure}
    \caption{Scatterplot of simulated data for sensitivity analysis.}
    \label{fig:sens-analysis-scatter}
  \end{figure}

\begin{figure}[ht]
    \centering
    \begin{subfigure}{0.45\linewidth}
      \includegraphics[width=\linewidth]{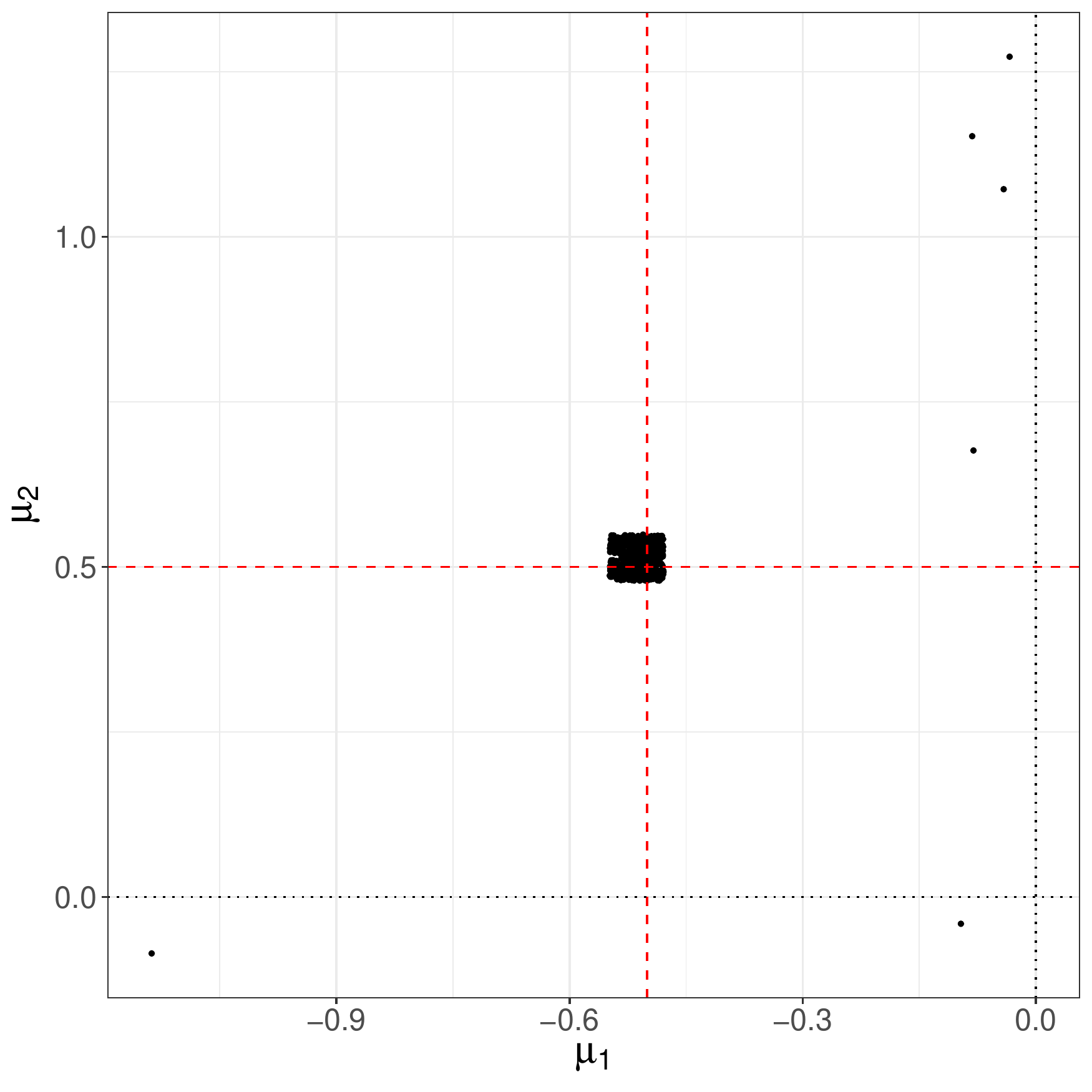}
      \caption{$\sqrt{p} \lambda_{x} = 10$, $\sigma_{1} = \sigma_{2} = 0.1$}
    \end{subfigure}
    \begin{subfigure}{0.45\linewidth}
      \includegraphics[width=\linewidth]{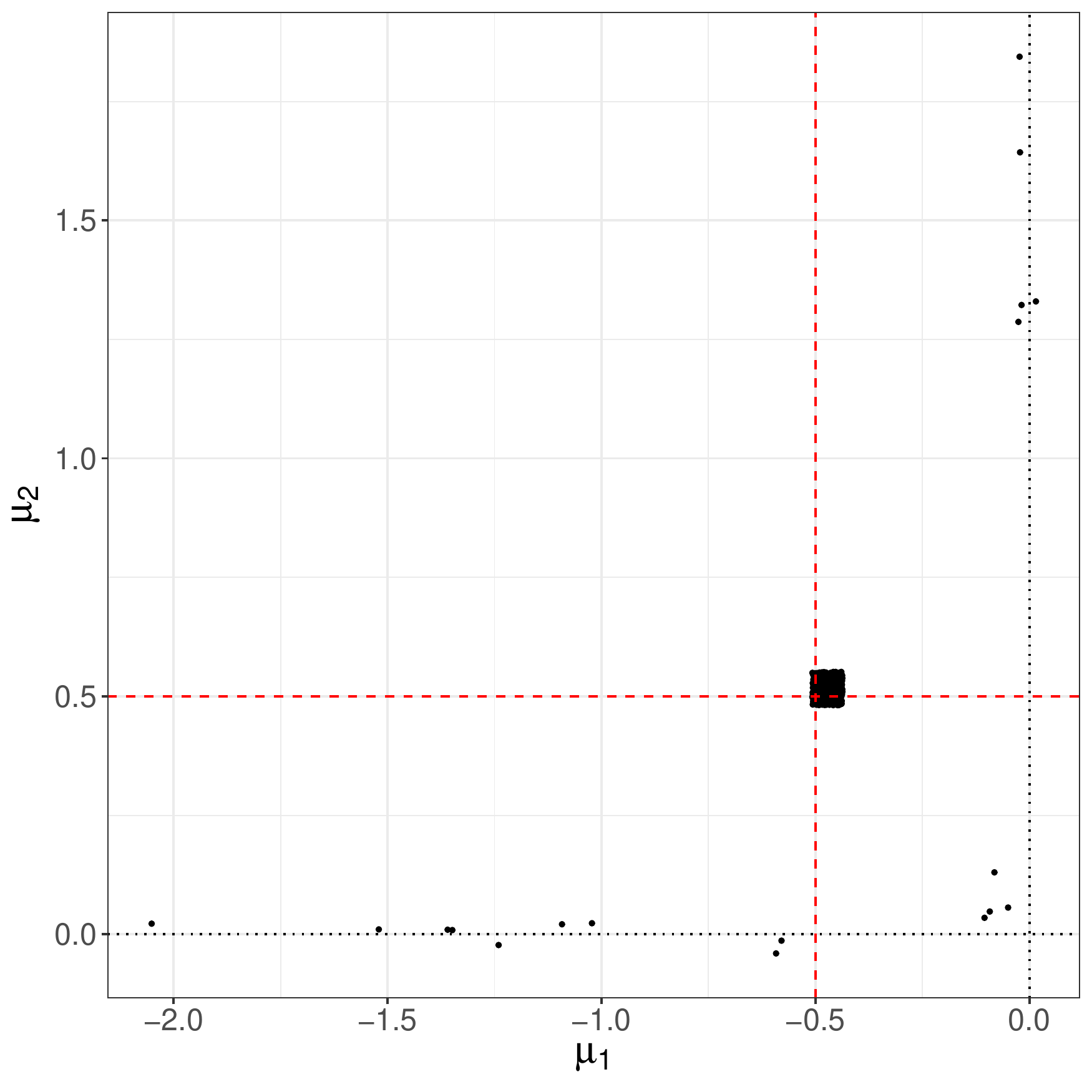}
      \caption{$\sqrt{p} \lambda_{x} = 5$, $\sigma_{1} = \sigma_{2} = 0.1$}
    \end{subfigure}
    \begin{subfigure}{0.45\linewidth}
      \includegraphics[width=\linewidth]{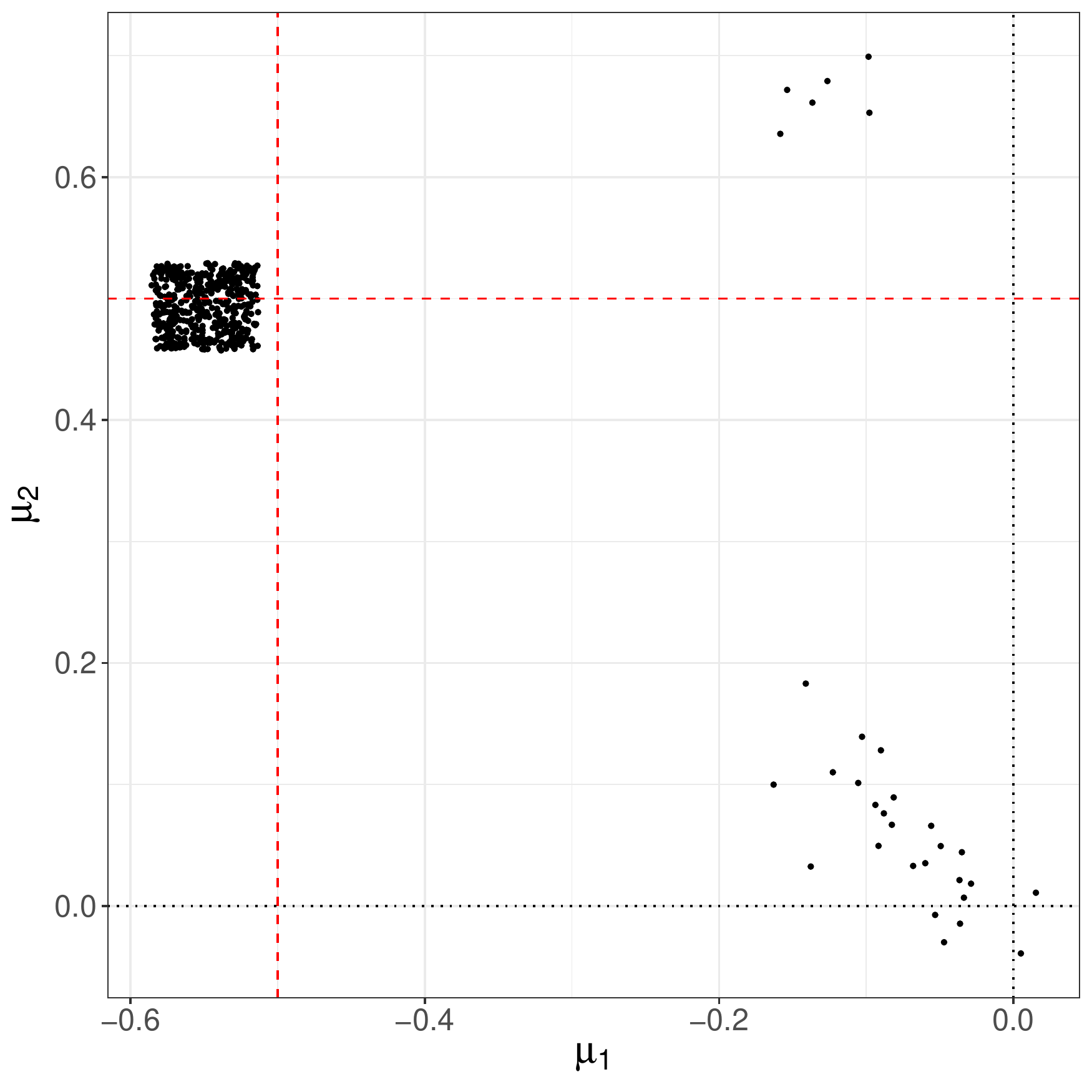}
      \caption{$\sqrt{p} \lambda_{x} = 10$, $\sigma_{1} = \sigma_{2} = 0.3$}
    \end{subfigure}
    \begin{subfigure}{0.45\linewidth}
      \includegraphics[width=\linewidth]{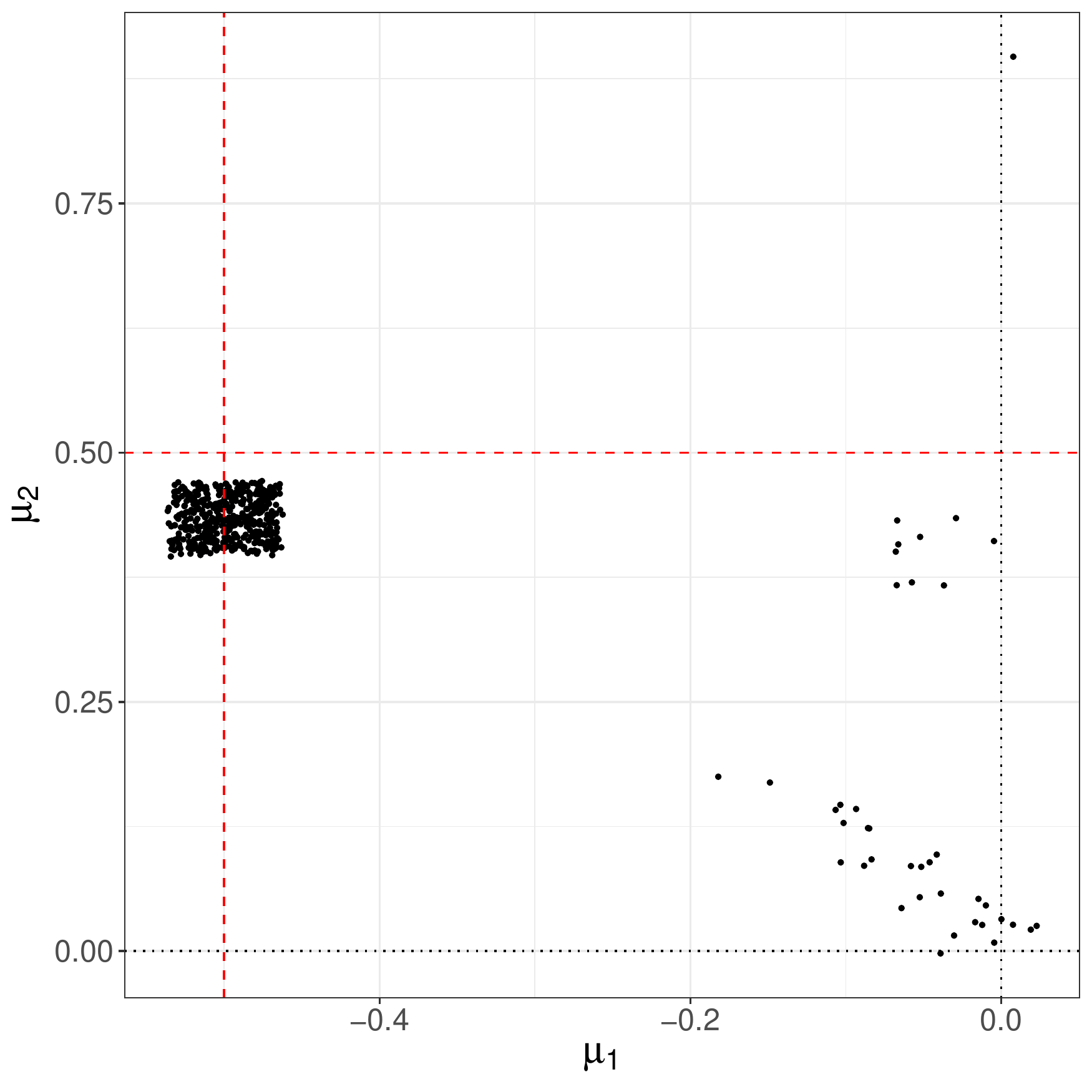}
      \caption{$\sqrt{p} \lambda_{x} = 5$, $\sigma_{1} = \sigma_{2} = 0.3$}
    \end{subfigure}
    \caption{Scatterplots of $\mu$ estimates from fitting MR-Path with 500 differrent
      starting values on simulated data with different $\sqrt{p} \lambda_{x}$ (columns)
      and $\sigma_{1} = \sigma_{2}$ (rows). True values of $\mu_{1}$ and $\mu_{2}$ are
      shown as red dashed lines.}
    \label{fig:sens-scatter}
  \end{figure}

  It is well known that the vanilla EM algorithm for Gaussian mixture models is
  sensitive to the initial values, especially when clusters overlap (\cite{biernacki2003choosing,shireman2017examining}).
  In this section, we conduct a small-scale simulation study to evaluate how sensitive our proposed
  MC-EM algorithm is to initial values. In this simulation study, we set $K = 2$,
  $\pi_{1} = \pi_{2}$, $\mu = (-0.5, 0.5)$.
  We vary the instrument strength by setting $\sqrt{p} \lambda_{x} = 5 \text{ or } 10$
  and the degree to which clusters overlap by setting $\sigma_{1} = \sigma_{2}$ to be
  either $0.1$ (low overlap) or $0.3$ (high overlap). We simulate data from our proposed
  model with these parameters (shown in \cref{fig:sens-analysis-scatter}) and apply the MC-EM algorithm with 500 different starting values.
  We plot the resulting $\mu$ estimates in \cref{fig:sens-scatter}. These preliminary
  results suggest that the MC-EM algorithm becomes more sensitive to starting values as
  the degree of cluster overlap increases. In the cases where
  $\sigma_{1} = \sigma_{2} = 0.1$, most of the estimates are close to the true values
  with a few estimates deviating from it. However, when $\sigma_{1} = \sigma_{2} = 0.3$,
  most of the estimates are slightly biased from the truth with a small cluster of
  estimates close to the origin.


\end{appendix}

\begin{acks}[Acknowledgments]
  We would like to thank Xuelu Wang for helpful comments on the type II diabetes
  example.
\end{acks}

\bibliographystyle{imsart-nameyear} 
\bibliography{references} 


\end{document}